\documentstyle[amsfonts,eqsecnum,preprint,aps]{revtex}
\begin{document}
\draft
\title{Polyakov's string classical mechanics.}
\author{Massimo Materassi}
\author{materassi@pg.infn.it, materassi@fi.infn.it}
\address{Department of Physics, University of Perugia (Italy)}
\date{\today }
\maketitle

\begin{abstract}

This paper is almost an exercise in which the Hamiltonian scheme is developed
for Polyakov's classical string, by following the usual framework suggested
by Dirac and Bergman for the reduction of gauge theories to their essential
physical degrees of freedom.
The results collected here will be useful in some forthcoming papers, where
strings will be studied in the unusual context of Wigner-covariant rest
frame theory.

After a short introduction outlining the work, the Lagrangean scheme is
presented in Section II, where the classical equivalence between Polyakov and
Nambu-Goto string is rederived. In Section III the Hamiltonian framework
is worked out, primary and secondary constraints are deduced. Then Lagrange
multipliers are introduced; finally the Hamiltonian equations of motion are
presented.

In Section IV gauge symmetries are treated, by constructing their canonical
generators; then some gauge degrees of freedom are eliminated by Dirac-Bergmann fixing procedure.
In this paper only the gauge-freedom coming from primary constraints is fixed,
while secondary constraints will need a deeper analysis. At the end of

Section IV the classical version of Virasoro algebra is singled out as that
of the secondary constraints, surviving to the first gauge fixing.
\end{abstract}

\bigskip

\narrowtext

\newpage  

\section{Introduction.}

In order to construct a good theory of relativistic insulated systems,
formed by interacting subsystems, we have chosen the framework of
Wigner-covariant rest frame scheme (see \cite{Lusanna.1}, \cite{lus.col.1}, 
\cite{lus.col.2}, \cite{LUSALBA}, \cite{matelonghi-kg}, \cite{matelonghi-bms}
\cite{matelusa-kg}, \cite{Bigazzi}, \cite{materassi}): in this context it is
possible to handle relativistic covariance in such a way that extended
relativistic systems (like many body systems and fields) can be treated in a
self consistent one-time formulation; many advantages of this approach are
known: preservation of covariance in an easily controlled way, an easier
form of the cluster decomposition property, the prospect to use almost
newtonian technologies. This line of research follows the steps suggested
many years ago by Dirac \cite{DIR}, when he introduced a space-like
foliation of Minkowski spacetime giving an invariant definition of
simultaneity.

Along this line of research many progresses have been achieved, some
Wigner-covariant rest frame theories have already been formulated: that of $%
N $ relativistic particles interacting with an electromagnetic field \cite
{Lusanna.1}, the one of charged particles interacting via Liendard-Wiechert
potentials in the abelian as well as Yang-Mills case \cite{LUSALBA}, that of
a free Dirac field. The next step we've been working on is the definition of
canonical center--of-mass vs relative variables for the fields \cite
{matelonghi-kg}, \cite{matelonghi-bms}, \cite{matelusa-kg}, \cite{materassi}%
, so that particles and radiation will be treated in the same way.

An interesting experiment will be the application of the same framework to
relativistic stringy objects, starting from classical Nambu-Goto string \cite
{matebiga}.

The motivation of this paper is essentially to review briefly the classical
mechanics of Polyakov's string, deriving its main features, by using
Dirac-Bergman approach for singular Lagrangean systems \cite{Teitelboim}.
The results here presented (which are not new, but are worked out explicitly
in a didactic feature) will be very useful in those forthcoming works about
Nambu-Goto string, whose dynamics is ''included'' in Polyakov's theory.

Polyakov's string classical theory is presented both in Lagrangean as well
as canonical terms; primary and secondary constraints are singled out, their
effects as canonical generators of classical gauge transformations are
studied.

The main aim in this paper is to stress the canonical mechanism of
generating Virasoro symmetries, in order to underline the background in
which one has to move when discussing Dirac-Bergman constraint reduction for
the classical string \cite{matebiga}.

With the simplest form of Polyakov's action (omitting the Liouville terms
suggested in \cite{marnelius} for sake of simplicity) the worldsheet metric $%
h_{{\rm ab}}\left( \sigma ,\tau \right) $ is pure gauge and it is possible
to gauge it away by using a suitable transformation generated by $\pi ^{{\rm %
ab}}\left( \sigma ,\tau \right) $, the worldsheet canonical momenta that are
the primary constraints of this theory. Stability conditions for $\pi ^{{\rm %
ab}}\left( \sigma ,\tau \right) $ leads to two secondary constraints, which
are the Virasoro constraints: here they are introduced as those symmetries
surviving after conformal gauge-fixing the metric on the worlsheet.

\section{Lagrangean framework.}

We'll work with a bosonic string described by {\it Polyakov's action} and
embedded in a Minkowskian flat spacetime $M_D$ of signature $\left(
1,D-1\right) $. The action is: 
\begin{equation}
S=-\frac T2\eta _{\mu \nu }%
%TCIMACRO{\dint }
%BeginExpansion
\displaystyle \int 
%EndExpansion
d\tau 
%TCIMACRO{\dint }
%BeginExpansion
\displaystyle \int 
%EndExpansion
d\sigma \sqrt{h}h^{{\rm ab}}\partial _{{\rm a}}X^\mu \partial _{{\rm b}%
}X^\nu .  \label{azione.pol}
\end{equation}
String motion is the evolution of the following Lagrangean coordinates 
\[
\begin{array}{cc}
X^\mu \left( \sigma \right) , & h^{{\rm ab}}\left( \sigma \right)
\end{array}
\]
with respect to the scalar parameter $\tau $.

Our point of view is to consider $X^\mu $ and $h^{{\rm ab}}$ as classical
fields in the curved $1+1$ background spanned by $\left( \sigma ,\tau
\right) $, referred to as {\it world sheet} ${\Bbb V}_2$: the freely chosen $%
{\Bbb V}_2$ coordinates will be scalars under spacetime Poincar\'e
transformation.

There exists a {\it worldsheet tensor calculus} derived from the $1+1$
metric $h^{{\rm ab}}$, with metric-adapted connections, covariant
derivatives, parallel transport and so on, while tensor calculus is trivial
for the spacetime (which is flat).

The only relationship between worldsheet geometry and spacetime is that in
order for string motion to have a causal coherence, the ${\Bbb V}_2$-tangent
vectors will be one spacelike and one timelike: 
\begin{equation}
\begin{array}{cc}
\eta _{\mu \nu }\partial _\tau X^\mu \partial _\tau X^\nu <0, & \eta _{\mu
\nu }\partial _\sigma X^\mu \partial _\sigma X^\nu >0,
\end{array}
\label{v.1}
\end{equation}
(the spacetime metric is $\eta =\left( +1,-1,-1,-1\right) $) and this allows
us to assign the conditions 
\begin{equation}
\begin{array}{cc}
X\left( \sigma ,\tau _0\right) =X_0\left( \sigma \right) , & \partial _\tau
X\left( \sigma ,\tau _0\right) =V_0\left( \sigma \right)
\end{array}
\label{buon.pdC}
\end{equation}
as well posed initial values.

We will refer to the following quantity 
\begin{equation}
L=-\frac T2\eta _{\mu \nu }%
%TCIMACRO{\dint }
%BeginExpansion
\displaystyle \int 
%EndExpansion
d\sigma \sqrt{h}h^{{\rm ab}}\partial _{{\rm a}}X^\mu \partial _{{\rm b}}X^\nu
\label{lag.1}
\end{equation}
as the {\it string Lagrangean}, which can be evaluated integrating the
Lagrangean linear density 
\begin{equation}
{\cal L}=-\frac T2\eta _{\mu \nu }\sqrt{h}h^{{\rm ab}}\partial _{{\rm a}%
}X^\mu \partial _{{\rm b}}X^\nu  \label{denslag.1}
\end{equation}
along the string.

With the positions 
\[
\begin{array}{cc}
\partial _\tau X=\dot X, & \partial _\sigma X=X^{\prime },
\end{array}
\]
the Lagrangean can be rewritten as 
\begin{equation}
L=-\frac T2\eta _{\mu \nu }%
%TCIMACRO{\dint }
%BeginExpansion
\displaystyle \int 
%EndExpansion
d\sigma \sqrt{h}\left( h^{\tau \tau }\dot X^\mu \dot X^\nu +2h^{\tau \sigma }%
\dot X^\mu X^{\prime \nu }+h^{\sigma \sigma }X^{\prime \mu }X^{\prime \nu
}\right) ,  \label{lag.2}
\end{equation}
while its linear density reads: 
\begin{equation}
{\cal L}=-\frac T2\eta _{\mu \nu }\sqrt{h}\left( h^{\tau \tau }\dot X^\mu 
\dot X^\nu +2h^{\tau \sigma }\dot X^\mu X^{\prime \nu }+h^{\sigma \sigma
}X^{\prime \mu }X^{\prime \nu }\right) .  \label{denslag.2}
\end{equation}
Since the worldsheet signature is $\left( 1,1\right) $ the factor $\sqrt{h}$
is the square root of 
\begin{equation}
h=-\det \left\| h_{{\rm ab}}\right\| .  \label{segn.h}
\end{equation}
From Binet theorem 
\begin{equation}
\det \left\| h^{{\rm ab}}\right\| =\frac 1{\det \left\| h_{{\rm ab}}\right\| 
}=-\frac 1h\Rightarrow h=\frac 1{\left( h^{\tau \sigma }\right) ^2-h^{\tau
\tau }h^{\sigma \sigma }},  \label{segn.h.1}
\end{equation}
and so: 
\begin{equation}
{\cal L}=-\frac{T\eta _{\mu \nu }}{2\sqrt{\left( h^{\tau \sigma }\right)
^2-h^{\tau \tau }h^{\sigma \sigma }}}\left( h^{\tau \tau }\dot X^\mu \dot X%
^\nu +2h^{\tau \sigma }\dot X^\mu X^{\prime \nu }+h^{\sigma \sigma
}X^{\prime \mu }X^{\prime \nu }\right) .  \label{denslag.2.bis}
\end{equation}
We will need this form for ${\cal L}$ later. We shall work out the
Lagrangean framework treating $X$ and $h$ as independent variables.

The metric will obey the following Lagrangean equations 
\begin{equation}
\begin{array}{cc}
\partial _{{\rm a}} 
%TCIMACRO{
%\dfrac{\partial {\cal L}}{\partial \left( \partial _{{\rm a}}X^\mu \right) } }
%BeginExpansion
{\displaystyle {\partial {\cal L} \over \partial \left( \partial _{{\rm a}}X^\mu \right) }}
%EndExpansion
-%
%TCIMACRO{\dfrac{\partial {\cal L}}{\partial X^\mu } }
%BeginExpansion
{\displaystyle {\partial {\cal L} \over \partial X^\mu }}
%EndExpansion
=0, & \partial _{{\rm a}} 
%TCIMACRO{
%\dfrac{\partial {\cal L}}{\partial \left( \partial _{{\rm a}}h^{{\rm bc}}\right) } }
%BeginExpansion
{\displaystyle {\partial {\cal L} \over \partial \left( \partial _{{\rm a}}h^{{\rm bc}}\right) }}
%EndExpansion
-%
%TCIMACRO{\dfrac{\partial {\cal L}}{\partial h^{{\rm bc}}} }
%BeginExpansion
{\displaystyle {\partial {\cal L} \over \partial h^{{\rm bc}}}}
%EndExpansion
=0,
\end{array}
\label{v.2}
\end{equation}
and from (\ref{denslag.1}) we get: 
\begin{equation}
\begin{array}{cc}
%TCIMACRO{
%\dfrac{\partial {\cal L}}{\partial \left( \partial _{{\rm a}}X^\mu \right) } }
%BeginExpansion
{\displaystyle {\partial {\cal L} \over \partial \left( \partial _{{\rm a}}X^\mu \right) }}
%EndExpansion
=-T\sqrt{h}\partial ^{{\rm a}}X_\mu , & 
%TCIMACRO{\dfrac{\partial {\cal L}}{\partial X^\mu } }
%BeginExpansion
{\displaystyle {\partial {\cal L} \over \partial X^\mu }}
%EndExpansion
=0,
\end{array}
\label{v.3}
\end{equation}
and even: 
\begin{equation}
\begin{array}{ccc}
\partial _{{\rm a}} 
%TCIMACRO{
%\dfrac{\partial {\cal L}}{\partial \left( \partial _{{\rm a}}h^{{\rm bc}}\right) } }
%BeginExpansion
{\displaystyle {\partial {\cal L} \over \partial \left( \partial _{{\rm a}}h^{{\rm bc}}\right) }}
%EndExpansion
=0, &  & 
%TCIMACRO{\dfrac{\partial {\cal L}}{\partial h^{{\rm bc}}} }
%BeginExpansion
{\displaystyle {\partial {\cal L} \over \partial h^{{\rm bc}}}}
%EndExpansion
=-%
%TCIMACRO{\dfrac{T\sqrt{h}}2 }
%BeginExpansion
{\displaystyle {T\sqrt{h} \over 2}}
%EndExpansion
\left( \partial _{{\rm b}}X^\mu \partial _{{\rm c}}X_\mu -%
%TCIMACRO{\dfrac{h_{{\rm bc}}}2 }
%BeginExpansion
{\displaystyle {h_{{\rm bc}} \over 2}}
%EndExpansion
h^{{\rm ef}}\partial _{{\rm e}}X^\mu \partial _{{\rm f}}X_\mu \right) ;
\end{array}
\label{v.4}
\end{equation}
So we will write: 
\[
\begin{array}{ccc}
\partial _{{\rm a}}\left( \sqrt{h}\partial ^{{\rm a}}X_\mu \right) =0, &  & 
\partial _{{\rm b}}X^\mu \partial _{{\rm c}}X_\mu -%
%TCIMACRO{\dfrac{h_{{\rm bc}}}2 }
%BeginExpansion
{\displaystyle {h_{{\rm bc}} \over 2}}
%EndExpansion
h^{{\rm ef}}\partial _{{\rm e}}X^\mu \partial _{{\rm f}}X_\mu =0.
\end{array}
\]
Let us explicit the first equation to get\footnote{%
Here the relationship 
\begin{equation}
%TCIMACRO{\dfrac \partial {\partial h^{{\rm cd}}} }
%BeginExpansion
{\displaystyle {\partial \over \partial h^{{\rm cd}}}}
%EndExpansion
\sqrt{h}=-\frac 12\sqrt{h}h_{{\rm cd}},  \label{derivdet}
\end{equation}
is usually employed.}: 
\begin{equation}
\begin{array}{ccc}
\partial _{{\rm a}}\partial ^{{\rm a}}X_\mu =%
%TCIMACRO{\dfrac 12 }
%BeginExpansion
{\displaystyle {1 \over 2}}
%EndExpansion
h_{{\rm cd}}\partial _{{\rm a}}h^{{\rm cd}}\partial ^{{\rm a}}X_\mu , &  & 
\partial _{{\rm b}}X^\mu \partial _{{\rm c}}X_\mu -%
%TCIMACRO{\dfrac{h_{{\rm bc}}}2 }
%BeginExpansion
{\displaystyle {h_{{\rm bc}} \over 2}}
%EndExpansion
h^{{\rm ef}}\partial _{{\rm e}}X^\mu \partial _{{\rm f}}X_\mu =0.
\end{array}
\label{v.5}
\end{equation}

First let us focus the relationship 
\begin{equation}
\partial _{{\rm b}}X^\mu \partial _{{\rm c}}X_\mu -%
%TCIMACRO{\dfrac{h_{{\rm bc}}}2 }
%BeginExpansion
{\displaystyle {h_{{\rm bc}} \over 2}}
%EndExpansion
h^{{\rm ef}}\partial _{{\rm e}}X^\mu \partial _{{\rm f}}X_\mu =0,
\label{v.6}
\end{equation}
which is simply Euler-Lagrange equation for the metric: with the position 
\begin{equation}
G_{{\rm bc}}=\partial _{{\rm b}}X^\mu \partial _{{\rm c}}X_\mu  \label{v.7}
\end{equation}
it reads: 
\[
G_{{\rm bc}}=%
%TCIMACRO{\dfrac{h_{{\rm bc}}}2}
%BeginExpansion
{\displaystyle {h_{{\rm bc}} \over 2}}
%EndExpansion
h^{{\rm ef}}G_{{\rm ef}}=%
%TCIMACRO{\dfrac{h_{{\rm bc}}}2}
%BeginExpansion
{\displaystyle {h_{{\rm bc}} \over 2}}
%EndExpansion
%TCIMACRO{\limfunc{tr}}
%BeginExpansion
\mathop{\rm tr}
%EndExpansion
\left\| G\right\| , 
\]
and from the equation of motion 
\begin{equation}
G_{{\rm bc}}=%
%TCIMACRO{\dfrac{h_{{\rm bc}}}2 }
%BeginExpansion
{\displaystyle {h_{{\rm bc}} \over 2}}
%EndExpansion
%TCIMACRO{\limfunc{tr} }
%BeginExpansion
\mathop{\rm tr}
%EndExpansion
\left\| G\right\|  \label{v.8}
\end{equation}
we get: 
\begin{equation}
\det \left\| G\right\| =%
%TCIMACRO{\dfrac{\left( \limfunc{tr}\left\| G\right\| \right) ^2}4 }
%BeginExpansion
{\displaystyle {\left( \mathop{\rm tr}\left\| G\right\| \right) ^2 \over 4}}
%EndExpansion
\det \left\| h\right\| .  \label{v.9}
\end{equation}
The absolute value of (\ref{v.9}) is 
\begin{equation}
G=%
%TCIMACRO{\dfrac{\left( \limfunc{tr}\left\| G\right\| \right) ^2}4 }
%BeginExpansion
{\displaystyle {\left( \mathop{\rm tr}\left\| G\right\| \right) ^2 \over 4}}
%EndExpansion
h,  \label{v.10}
\end{equation}
which becomes 
\begin{equation}
\sqrt{G}=%
%TCIMACRO{\dfrac{\limfunc{tr}\left\| G\right\| }2 }
%BeginExpansion
{\displaystyle {\mathop{\rm tr}\left\| G\right\|  \over 2}}
%EndExpansion
\sqrt{h},  \label{v.11}
\end{equation}
and writing $%
%TCIMACRO{\limfunc{tr}}
%BeginExpansion
\mathop{\rm tr}
%EndExpansion
\left\| G\right\| $ in more explicit form as in (\ref{v.7}) 
\begin{equation}
%TCIMACRO{\limfunc{tr} }
%BeginExpansion
\mathop{\rm tr}
%EndExpansion
\left\| G\right\| =h^{{\rm ab}}\partial _{{\rm a}}X^\mu \partial _{{\rm b}%
}X_\mu  \label{v.12}
\end{equation}
one recognizes: 
\begin{equation}
\sqrt{G}=%
%TCIMACRO{\dfrac{\sqrt{h}}2 }
%BeginExpansion
{\displaystyle {\sqrt{h} \over 2}}
%EndExpansion
h^{{\rm ab}}\partial _{{\rm a}}X^\mu \partial _{{\rm b}}X_\mu .  \label{v.13}
\end{equation}

So Euler-Lagrange equation for the independent variable metric $h^{{\rm ab}}$
reads: 
\begin{equation}
\begin{array}{ccc}
%TCIMACRO{\dfrac{\sqrt{h}}2 }
%BeginExpansion
{\displaystyle {\sqrt{h} \over 2}}
%EndExpansion
h^{{\rm ab}}\partial _{{\rm a}}X^\mu \partial _{{\rm b}}X_\mu =\sqrt{G}, & 
& G_{{\rm bc}}=\partial _{{\rm b}}X^\mu \partial _{{\rm c}}X_\mu ,
\end{array}
\label{moto.h}
\end{equation}
which translates (\ref{azione.pol}) into plain Nambu-Goto action: 
\begin{equation}
S=-T%
%TCIMACRO{\dint }
%BeginExpansion
\displaystyle \int 
%EndExpansion
d\tau 
%TCIMACRO{\dint }
%BeginExpansion
\displaystyle \int 
%EndExpansion
d\sigma \sqrt{G}.  \label{azione.ng.1}
\end{equation}
{\it Nambu-Goto action is obtained from Polyakov's (\ref{azione.pol}) using
Euler-Lagrange equations for the worldsheet metric.}

Action (\ref{azione.ng.1}) does really be Nambu-Goto action, i.e. $-T$ times
the measure of ${\Bbb V}_2$, considering 
\begin{equation}
%TCIMACRO{\limfunc{mis} }
%BeginExpansion
\mathop{\rm mis}
%EndExpansion
{\Bbb V}_2=%
%TCIMACRO{\dint }
%BeginExpansion
\displaystyle \int 
%EndExpansion
d\tau 
%TCIMACRO{\dint }
%BeginExpansion
\displaystyle \int 
%EndExpansion
d\sigma \sqrt{G},  \label{v.14}
\end{equation}
that is just considering sheet-tensor $G_{{\rm bc}}=\partial _{{\rm b}}X^\mu
\partial _{{\rm c}}X_\mu $ as the worldsheet metric. So we have to choose
the option: 
\begin{equation}
\begin{array}{ccc}
h_{{\rm bc}}=G_{{\rm bc}} & \Rightarrow & h_{{\rm bc}}=\eta _{\mu \nu
}\partial _{{\rm b}}X^\mu \partial _{{\rm c}}X^\nu .
\end{array}
\label{embedh}
\end{equation}
Equation (\ref{embedh}) simply embeds ${\Bbb V}_2$ into $M_D$. One can find 
\cite{Hatfield} very simply the generalization to General relativistic free
falling string, moving on a rigid background $g_{\mu \nu }\left( X\right) $.

When written explicitly, Nambu-Goto action reads 
\begin{equation}
S_{%
%TCIMACRO{\limfunc{NG} }
%BeginExpansion
\mathop{\rm NG}
%EndExpansion
}=-T%
%TCIMACRO{\dint }
%BeginExpansion
\displaystyle \int 
%EndExpansion
d\tau 
%TCIMACRO{\dint }
%BeginExpansion
\displaystyle \int 
%EndExpansion
d\sigma \sqrt{\left| \dot X^\mu \dot X_\mu X^{\prime \alpha }X_\alpha
^{\prime }-\left( \dot X^\mu X_\mu ^{\prime }\right) \right| },
\label{azione.ng.2}
\end{equation}
and so Lagrangean linear density is: 
\begin{equation}
{\cal L}_{%
%TCIMACRO{\limfunc{NG} }
%BeginExpansion
\mathop{\rm NG}
%EndExpansion
}=-T\sqrt{\left| \dot X^\mu \dot X_\mu X^{\prime \alpha }X_\alpha ^{\prime
}-\left( \dot X^\mu X_\mu ^{\prime }\right) \right| }.  \label{denslag.3}
\end{equation}

The fact that Nambu-Goto theory is included in Polyakov's one will let us
use the content of this paper in \cite{matebiga}.

Finally, let's consider Euler-Lagrange equations of motion for the string
variables: 
\begin{equation}
\partial _{{\rm a}}\partial ^{{\rm a}}X_\mu =%
%TCIMACRO{\dfrac 12 }
%BeginExpansion
{\displaystyle {1 \over 2}}
%EndExpansion
h_{{\rm cd}}\partial _{{\rm a}}h^{{\rm cd}}\partial ^{{\rm a}}X_\mu .
\label{v.17}
\end{equation}

\section{Hamiltonian framework.}

Let's organize an {\it Hamiltonian scheme for the classical free string},
described by (\ref{azione.pol}), whose Lagrangean linear density is: 
\begin{equation}
{\cal L}=-\frac{T\eta _{\mu \nu }}{2\sqrt{\left( h^{\tau \sigma }\right)
^2-h^{\tau \tau }h^{\sigma \sigma }}}\left( h^{\tau \tau }\dot X^\mu \dot X%
^\nu +2h^{\tau \sigma }\dot X^\mu X^{\prime \nu }+h^{\sigma \sigma
}X^{\prime \mu }X^{\prime \nu }\right) .  \label{denslag.4}
\end{equation}
The $X$ and the $h$ variables will be thought of as independent.

\subsection{Primary constraints and ''constraintless'' Hamiltonian.}

To realize a canonical version of the theory described by (\ref{denslag.4})
we'll have to evaluate canonical momenta of string variables 
\begin{equation}
P_\mu =\frac{\partial {\cal L}}{\partial \dot X^\mu }  \label{momento.x}
\end{equation}
as well as of sheet variables 
\begin{equation}
\pi _{{\rm ab}}=\frac{\partial {\cal L}}{\partial \dot h^{{\rm ab}}},
\label{momento.h}
\end{equation}
assuming equal time Poisson brackets: 
\begin{equation}
\left\{ 
\begin{array}{l}
\left\{ X^\mu \left( \sigma ,\tau \right) ,P_\nu \left( \sigma ^{\prime
},\tau \right) \right\} =\eta _\nu ^\mu \delta \left( \sigma -\sigma
^{\prime }\right) , \\ 
\\ 
\left\{ h^{{\rm ab}}\left( \sigma ,\tau \right) ,\pi _{{\rm cd}}\left(
\sigma ^{\prime },\tau \right) \right\} =%
%TCIMACRO{\dfrac 12 }
%BeginExpansion
{\displaystyle {1 \over 2}}
%EndExpansion
\left( \delta _{{\rm c}}^{{\rm a}}\delta _{{\rm d}}^{{\rm b}}+\delta _{{\rm d%
}}^{{\rm a}}\delta _{{\rm c}}^{{\rm b}}\right) \delta \left( \sigma -\sigma
^{\prime }\right) , \\ 
\\ 
\left\{ X^\mu \left( \sigma ,\tau \right) ,X^\nu \left( \sigma ,\tau \right)
\right\} =\left\{ P_\mu \left( \sigma ,\tau \right) ,P_\nu \left( \sigma
^{\prime },\tau \right) \right\} =0, \\ 
\\ 
\left\{ h^{{\rm ab}}\left( \sigma ,\tau \right) ,h^{{\rm cd}}\left( \sigma
^{\prime },\tau \right) \right\} =\left\{ \pi _{{\rm ab}}\left( \sigma ,\tau
\right) ,\pi _{{\rm cd}}\left( \sigma ^{\prime },\tau \right) \right\} =0.
\end{array}
\right.  \label{v.19}
\end{equation}

Since 
\begin{equation}
%TCIMACRO{
%\dfrac{\partial {\cal L}}{\partial \left( \partial _{{\rm a}}h^{{\rm bc}}\right) } }
%BeginExpansion
{\displaystyle {\partial {\cal L} \over \partial \left( \partial _{{\rm a}}h^{{\rm bc}}\right) }}
%EndExpansion
=0  \label{v.20}
\end{equation}
canonical momenta $\pi _{{\rm ab}}$ vanish identically, so the theory from (%
\ref{denslag.4}) is a constrained one, with primary constraints: 
\begin{equation}
\pi _{{\rm ab}}\approx 0.  \label{vincprim}
\end{equation}
Due to the symmetry 
\[
h^{{\rm bc}}=h^{{\rm cb}} 
\]
the independent primary constraints are of course only three, for example
the following ones: 
\begin{equation}
\begin{array}{ccccc}
\pi _{\tau \tau }\approx 0, &  & \pi _{\tau \sigma }\approx 0, &  & \pi
_{\sigma \sigma }\approx 0,
\end{array}
\label{v.21}
\end{equation}
and form a first class set, as prescribed by (\ref{v.19}). These constraints
are {\it strongly} first class, because of assumptions (\ref{v.19}). We'll
first build up the ''constraintless'' part of the Hamiltonian, then we'll
include constraints with Lagrange multipliers.

String variables $X$ have nonvanishing momenta: 
\begin{equation}
P_\mu =-T\sqrt{h}\left( h^{\tau \tau }\dot X_\mu +h^{\tau \sigma }X_\mu
^{\prime }\right) .  \label{v.22}
\end{equation}
From (\ref{v.22}) one can read back $\dot X_\mu $ in terms of $P_\mu $,
obtaining: 
\begin{equation}
\dot X_\mu =-\frac 1{Th^{\tau \tau }\sqrt{h}}P_\mu -\frac{h^{\tau \sigma }}{%
h^{\tau \tau }}X_\mu ^{\prime }.  \label{v.23}
\end{equation}
With (\ref{v.22}) and (\ref{v.23}) we're ready to write down the
''constraintless'' Hamiltonian density: 
\begin{equation}
{\cal H}_0=-\frac{\sqrt{\left( h^{\tau \sigma }\right) ^2-h^{\sigma \sigma
}h^{\tau \tau }}}{2Th^{\tau \tau }}P^\mu P_\mu -%
%TCIMACRO{\dfrac{h^{\tau \sigma }}{h^{\tau \tau }} }
%BeginExpansion
{\displaystyle {h^{\tau \sigma } \over h^{\tau \tau }}}
%EndExpansion
P^\mu X_\mu ^{\prime }-%
%TCIMACRO{\dfrac T2 }
%BeginExpansion
{\displaystyle {T \over 2}}
%EndExpansion
\frac{\sqrt{\left( h^{\tau \sigma }\right) ^2-h^{\sigma \sigma }h^{\tau \tau
}}}{h^{\tau \tau }}X^{\prime \mu }X_\mu ^{\prime }.  \label{densham.can}
\end{equation}
The whole ''constraintless'' Hamiltonian is obtained integrating ${\cal H}_0$
along the string, and it reads: 
\begin{equation}
H_0=-%
%TCIMACRO{\dint }
%BeginExpansion
\displaystyle \int 
%EndExpansion
\limits_0^\pi d\sigma \left( \frac{\sqrt{\left( h^{\tau \sigma }\right)
^2-h^{\sigma \sigma }h^{\tau \tau }}}{2Th^{\tau \tau }}P^\mu P_\mu +%
%TCIMACRO{\dfrac{h^{\tau \sigma }}{h^{\tau \tau }} }
%BeginExpansion
{\displaystyle {h^{\tau \sigma } \over h^{\tau \tau }}}
%EndExpansion
P^\mu X_\mu ^{\prime }+%
%TCIMACRO{\dfrac T2 }
%BeginExpansion
{\displaystyle {T \over 2}}
%EndExpansion
\frac{\sqrt{\left( h^{\tau \sigma }\right) ^2-h^{\sigma \sigma }h^{\tau \tau
}}}{h^{\tau \tau }}X^{\prime \mu }X_\mu ^{\prime }\right)  \label{ham.can}
\end{equation}
(let's remember Equation (\ref{segn.h.1})).

\subsection{Secondary constraints.}

Let's work out stability conditions for our primary constraints (\ref
{vincprim}). We could use the equations 
\[
\left\{ \pi _{{\rm ab}},H_{{\rm C}}\right\} \approx 0 
\]
(where $H_{{\rm C}}$ is the canonical Hamiltonian obtained from $H_0$ adding
a linear combination of primary constraints (\ref{v.21})), but there exist
an {\it easier method}\footnote{%
This method is reliable only if one assumes lagrangian and hamiltonian
motions completely equivalent.} which can be employed when some constraints
are canonical momenta: from Euler-Lagrange equations 
\[
\partial _{{\rm a}}%
%TCIMACRO{
%\dfrac{\partial {\cal L}}{\partial \left( \partial _{{\rm a}}h^{{\rm bc}}\right) }}
%BeginExpansion
{\displaystyle {\partial {\cal L} \over \partial \left( \partial _{{\rm a}}h^{{\rm bc}}\right) }}
%EndExpansion
-%
%TCIMACRO{\dfrac{\partial {\cal L}}{\partial h^{{\rm bc}}}}
%BeginExpansion
{\displaystyle {\partial {\cal L} \over \partial h^{{\rm bc}}}}
%EndExpansion
=0 
\]
one can see that canonical momenta obey 
\begin{equation}
\dot \pi _{{\rm ab}}=%
%TCIMACRO{\dfrac{\partial {\cal L}}{\partial h^{{\rm ab}}} }
%BeginExpansion
{\displaystyle {\partial {\cal L} \over \partial h^{{\rm ab}}}}
%EndExpansion
-\partial _\sigma 
%TCIMACRO{
%\dfrac{\partial {\cal L}}{\partial \left( \partial _\sigma h^{{\rm ab}}\right) } }
%BeginExpansion
{\displaystyle {\partial {\cal L} \over \partial \left( \partial _\sigma h^{{\rm ab}}\right) }}
%EndExpansion
,  \label{v.26}
\end{equation}
which is 
\begin{equation}
\dot \pi _{{\rm ab}}=%
%TCIMACRO{\dfrac{\partial {\cal L}}{\partial h^{{\rm ab}}} }
%BeginExpansion
{\displaystyle {\partial {\cal L} \over \partial h^{{\rm ab}}}}
%EndExpansion
\label{v.27}
\end{equation}
for us, due to Equation (\ref{v.20}). This is why we read stability
conditions of (\ref{v.21}) from the equations: 
\begin{equation}
\begin{array}{ccccc}
%TCIMACRO{\dfrac{\partial {\cal L}}{\partial h^{\tau \tau }} }
%BeginExpansion
{\displaystyle {\partial {\cal L} \over \partial h^{\tau \tau }}}
%EndExpansion
\approx 0, &  & 
%TCIMACRO{\dfrac{\partial {\cal L}}{\partial h^{\tau \sigma }} }
%BeginExpansion
{\displaystyle {\partial {\cal L} \over \partial h^{\tau \sigma }}}
%EndExpansion
\approx 0, &  & 
%TCIMACRO{\dfrac{\partial {\cal L}}{\partial h^{\sigma \sigma }} }
%BeginExpansion
{\displaystyle {\partial {\cal L} \over \partial h^{\sigma \sigma }}}
%EndExpansion
\approx 0.
\end{array}
\label{v.28}
\end{equation}

Remembering Equation (\ref{v.4}) 
\[
%TCIMACRO{\dfrac{\partial {\cal L}}{\partial h^{{\rm bc}}}}
%BeginExpansion
{\displaystyle {\partial {\cal L} \over \partial h^{{\rm bc}}}}
%EndExpansion
=-%
%TCIMACRO{\dfrac{T\sqrt{h}}2}
%BeginExpansion
{\displaystyle {T\sqrt{h} \over 2}}
%EndExpansion
\left( \partial _{{\rm b}}X^\mu \partial _{{\rm c}}X_\mu -%
%TCIMACRO{\dfrac{h_{{\rm bc}}}2}
%BeginExpansion
{\displaystyle {h_{{\rm bc}} \over 2}}
%EndExpansion
h^{{\rm ef}}\partial _{{\rm e}}X^\mu \partial _{{\rm f}}X_\mu \right) , 
\]
we translate conditions (\ref{v.21}) into the following system: 
\begin{equation}
\left\{ 
\begin{array}{l}
\dot X^\mu \dot X_\mu -%
%TCIMACRO{\dfrac{h_{\tau \tau }}2 }
%BeginExpansion
{\displaystyle {h_{\tau \tau } \over 2}}
%EndExpansion
h^{{\rm ef}}\partial _{{\rm e}}X^\mu \partial _{{\rm f}}X_\mu \approx 0, \\ 
\\ 
\dot X^\mu X_\mu ^{\prime }-%
%TCIMACRO{\dfrac{h_{\tau \sigma }}2 }
%BeginExpansion
{\displaystyle {h_{\tau \sigma } \over 2}}
%EndExpansion
h^{{\rm ef}}\partial _{{\rm e}}X^\mu \partial _{{\rm f}}X_\mu \approx 0, \\ 
\\ 
X^{\prime \mu }X_\mu ^{\prime }-%
%TCIMACRO{\dfrac{h_{\sigma \sigma }}2 }
%BeginExpansion
{\displaystyle {h_{\sigma \sigma } \over 2}}
%EndExpansion
h^{{\rm ef}}\partial _{{\rm e}}X^\mu \partial _{{\rm f}}X_\mu \approx 0.
\end{array}
\right.  \label{v.29}
\end{equation}
We have to use the expressions for $\dot X^\mu $ in terms of $P^\mu $ and
those of inverted metric $h_{{\rm ab}}$ in terms of $h^{{\rm ef}}$.

The inversion of the matrix 
\[
h^{{\rm ab}}=\left( 
\begin{array}{cc}
h^{\tau \tau } & h^{\tau \sigma } \\ 
h^{\tau \sigma } & h^{\sigma \sigma }
\end{array}
\right) 
\]
leads to 
\begin{equation}
h_{{\rm ab}}=\left( 
\begin{array}{cc}
-hh^{\sigma \sigma } & hh^{\tau \sigma } \\ 
hh^{\tau \sigma } & -hh^{\tau \tau }
\end{array}
\right) ,  \label{v.30}
\end{equation}
i.e.: 
\begin{equation}
\begin{array}{ccc}
h_{\tau \tau }=-hh^{\sigma \sigma }, & h_{\tau \sigma }=hh^{\tau \sigma }, & 
h_{\sigma \sigma }=-hh^{\tau \tau }.
\end{array}
\label{invert.h}
\end{equation}
So we get: 
\begin{equation}
\left\{ 
\begin{array}{l}
\dot X^\mu \dot X_\mu +%
%TCIMACRO{\dfrac{hh^{\sigma \sigma }}2 }
%BeginExpansion
{\displaystyle {hh^{\sigma \sigma } \over 2}}
%EndExpansion
h^{{\rm ef}}\partial _{{\rm e}}X^\mu \partial _{{\rm f}}X_\mu \approx 0, \\ 
\\ 
\dot X^\mu X_\mu ^{\prime }-%
%TCIMACRO{\dfrac{hh^{\tau \sigma }}2 }
%BeginExpansion
{\displaystyle {hh^{\tau \sigma } \over 2}}
%EndExpansion
h^{{\rm ef}}\partial _{{\rm e}}X^\mu \partial _{{\rm f}}X_\mu \approx 0, \\ 
\\ 
X^{\prime \mu }X_\mu ^{\prime }+%
%TCIMACRO{\dfrac{hh^{\tau \tau }}2 }
%BeginExpansion
{\displaystyle {hh^{\tau \tau } \over 2}}
%EndExpansion
h^{{\rm ef}}\partial _{{\rm e}}X^\mu \partial _{{\rm f}}X_\mu \approx 0.
\end{array}
\right.  \label{v.31}
\end{equation}

These constraints are not completely independent, they must be reduced to a
set of two independent ones only. In order to do this, we'll employ the
trace $h^{{\rm ef}}\partial _{{\rm e}}X^\mu \partial _{{\rm f}}X_\mu $ in
the {\it r\^ole} of a parameter, obtaining a system which is equivalent to (%
\ref{v.31}) but simpler than it, coupling together the equations two by two.
From the first equation we have $h^{{\rm ef}}\partial _{{\rm e}}X^\mu
\partial _{{\rm f}}X_\mu $ as 
\begin{equation}
h^{{\rm ef}}\partial _{{\rm e}}X^\mu \partial _{{\rm f}}X_\mu \approx -\frac 
2{hh^{\sigma \sigma }}\dot X^\mu \dot X_\mu  \label{traccia.2}
\end{equation}
and putting it into the second one\footnote{%
We work like this in order to never divide by the off-diagonal component $%
h^{\tau \sigma }$, since by its symmetry the metric tensor is point-by-point
diagonalizable, and it must be possible to put $h^{\tau \sigma }=0$ as {\it %
gauge fixing} without any unpleasant divergency.}: 
\begin{equation}
\dot X^\mu X_\mu ^{\prime }+\frac{h^{\tau \sigma }}{h^{\sigma \sigma }}\dot X%
^\mu \dot X_\mu \approx 0  \label{vincsec.1}
\end{equation}
Involving only coordinates and canonical momenta, it reads: 
\begin{equation}
\left( 
%TCIMACRO{
%\dfrac{2\left( h^{\tau \sigma }\right) ^2}{Th^{\sigma \sigma }\left( h^{\tau \tau }\right) ^2\sqrt{h}} }
%BeginExpansion
{\displaystyle {2\left( h^{\tau \sigma }\right) ^2 \over Th^{\sigma \sigma }\left( h^{\tau \tau }\right) ^2\sqrt{h}}}
%EndExpansion
-%
%TCIMACRO{\dfrac 1{Th^{\tau \tau }\sqrt{h}} }
%BeginExpansion
{\displaystyle {1 \over Th^{\tau \tau }\sqrt{h}}}
%EndExpansion
\right) P_\mu X^{\prime \mu }+\left( 
%TCIMACRO{
%\dfrac{\left( h^{\tau \sigma }\right) ^3}{h^{\sigma \sigma }\left( h^{\tau \tau }\right) ^2} }
%BeginExpansion
{\displaystyle {\left( h^{\tau \sigma }\right) ^3 \over h^{\sigma \sigma }\left( h^{\tau \tau }\right) ^2}}
%EndExpansion
-%
%TCIMACRO{\dfrac{h^{\tau \sigma }}{h^{\tau \tau }} }
%BeginExpansion
{\displaystyle {h^{\tau \sigma } \over h^{\tau \tau }}}
%EndExpansion
\right) X_\mu ^{\prime }X^{\prime \mu }+%
%TCIMACRO{\dfrac{h^{\tau \sigma }}{h^{\sigma \sigma }} }
%BeginExpansion
{\displaystyle {h^{\tau \sigma } \over h^{\sigma \sigma }}}
%EndExpansion
%TCIMACRO{\dfrac{P^\mu P_\mu }{T^2\left( h^{\tau \tau }\right) ^2h} }
%BeginExpansion
{\displaystyle {P^\mu P_\mu  \over T^2\left( h^{\tau \tau }\right) ^2h}}
%EndExpansion
\approx 0.  \label{vincsec.a}
\end{equation}

Let's put together first and third equation in (\ref{v.31}): we get $h^{{\rm %
ef}}\partial _{{\rm e}}X^\mu \partial _{{\rm f}}X_\mu $ from the first (as
in (\ref{traccia.2})) and put it into the third 
\begin{equation}
X^{\prime \mu }X_\mu ^{\prime }-\frac{h^{\tau \tau }}{h^{\sigma \sigma }}%
\dot X^\mu \dot X_\mu \approx 0,  \label{vincsec.2}
\end{equation}
getting something which becomes 
\begin{equation}
\left( 1- 
%TCIMACRO{
%\dfrac{\left( h^{\tau \sigma }\right) ^2}{h^{\sigma \sigma }h^{\tau \tau }} }
%BeginExpansion
{\displaystyle {\left( h^{\tau \sigma }\right) ^2 \over h^{\sigma \sigma }h^{\tau \tau }}}
%EndExpansion
\right) X^{\prime \mu }X_\mu ^{\prime }-%
%TCIMACRO{\dfrac 1{T^2h^{\sigma \sigma }h^{\tau \tau }h} }
%BeginExpansion
{\displaystyle {1 \over T^2h^{\sigma \sigma }h^{\tau \tau }h}}
%EndExpansion
P^\mu P_\mu - 
%TCIMACRO{
%\dfrac{2h^{\tau \sigma }}{Th^{\sigma \sigma }h^{\tau \tau }\sqrt{h}} }
%BeginExpansion
{\displaystyle {2h^{\tau \sigma } \over Th^{\sigma \sigma }h^{\tau \tau }\sqrt{h}}}
%EndExpansion
P^\mu X_\mu ^{\prime }\approx 0  \label{vincsec.b}
\end{equation}
considering (\ref{v.23}).

Finally, let's put together the second and the third equations in (\ref{v.31}%
): now $h^{{\rm ef}}\partial _{{\rm e}}X^\mu \partial _{{\rm f}}X_\mu $
comes from the third equation\footnote{%
...always in order to never have $h^{\tau \sigma }$ as a divider!} 
\begin{equation}
h^{{\rm ef}}\partial _{{\rm e}}X^\mu \partial _{{\rm f}}X_\mu \approx -\frac 
2{hh^{\tau \tau }}X^{\prime \mu }X_\mu ^{\prime }  \label{traccia.3}
\end{equation}
and it is put into the second one 
\begin{equation}
\dot X^\mu X_\mu ^{\prime }+\frac{h^{\tau \sigma }}{h^{\tau \tau }}X^{\prime
\mu }X_\mu ^{\prime }\approx 0,  \label{vincsec.3}
\end{equation}
which is translated into an expression involving only coordinates and
momenta: 
\begin{equation}
-\frac 1{Th^{\tau \tau }\sqrt{h}}P_\mu X^{\prime \mu }\approx 0.
\label{vincsec.c}
\end{equation}

Now we have only to put together constraints (\ref{vincsec.a}), (\ref
{vincsec.b}) and (\ref{vincsec.c}) to get secondary constraints ensuring
stability of the primary ones: 
\begin{equation}
\left\{ 
\begin{array}{l}
\left( 
%TCIMACRO{
%\dfrac{2\left( h^{\tau \sigma }\right) ^2}{Th^{\sigma \sigma }\left( h^{\tau \tau }\right) ^2\sqrt{h}} }
%BeginExpansion
{\displaystyle {2\left( h^{\tau \sigma }\right) ^2 \over Th^{\sigma \sigma }\left( h^{\tau \tau }\right) ^2\sqrt{h}}}
%EndExpansion
-%
%TCIMACRO{\dfrac 1{Th^{\tau \tau }\sqrt{h}} }
%BeginExpansion
{\displaystyle {1 \over Th^{\tau \tau }\sqrt{h}}}
%EndExpansion
\right) P_\mu X^{\prime \mu }+\left( 
%TCIMACRO{
%\dfrac{\left( h^{\tau \sigma }\right) ^3}{h^{\sigma \sigma }\left( h^{\tau \tau }\right) ^2} }
%BeginExpansion
{\displaystyle {\left( h^{\tau \sigma }\right) ^3 \over h^{\sigma \sigma }\left( h^{\tau \tau }\right) ^2}}
%EndExpansion
-%
%TCIMACRO{\dfrac{h^{\tau \sigma }}{h^{\tau \tau }} }
%BeginExpansion
{\displaystyle {h^{\tau \sigma } \over h^{\tau \tau }}}
%EndExpansion
\right) X_\mu ^{\prime }X^{\prime \mu }+%
%TCIMACRO{\dfrac{h^{\tau \sigma }}{h^{\sigma \sigma }} }
%BeginExpansion
{\displaystyle {h^{\tau \sigma } \over h^{\sigma \sigma }}}
%EndExpansion
%TCIMACRO{\dfrac{P^\mu P_\mu }{T^2\left( h^{\tau \tau }\right) ^2h} }
%BeginExpansion
{\displaystyle {P^\mu P_\mu  \over T^2\left( h^{\tau \tau }\right) ^2h}}
%EndExpansion
\approx 0, \\ 
\\ 
\left( 1- 
%TCIMACRO{
%\dfrac{\left( h^{\tau \sigma }\right) ^2}{h^{\sigma \sigma }h^{\tau \tau }} }
%BeginExpansion
{\displaystyle {\left( h^{\tau \sigma }\right) ^2 \over h^{\sigma \sigma }h^{\tau \tau }}}
%EndExpansion
\right) X^{\prime \mu }X_\mu ^{\prime }-%
%TCIMACRO{\dfrac 1{T^2h^{\sigma \sigma }h^{\tau \tau }h} }
%BeginExpansion
{\displaystyle {1 \over T^2h^{\sigma \sigma }h^{\tau \tau }h}}
%EndExpansion
P^\mu P_\mu - 
%TCIMACRO{
%\dfrac{2h^{\tau \sigma }}{Th^{\sigma \sigma }h^{\tau \tau }\sqrt{h}} }
%BeginExpansion
{\displaystyle {2h^{\tau \sigma } \over Th^{\sigma \sigma }h^{\tau \tau }\sqrt{h}}}
%EndExpansion
P^\mu X_\mu ^{\prime }\approx 0, \\ 
\\ 
-%
%TCIMACRO{\dfrac 1{Th^{\tau \tau }\sqrt{h}} }
%BeginExpansion
{\displaystyle {1 \over Th^{\tau \tau }\sqrt{h}}}
%EndExpansion
P_\mu X^{\prime \mu }\approx 0.
\end{array}
\right.  \label{vincsec.tot.1}
\end{equation}
The third one, which reads equally $P_\mu X^{\prime \mu }\approx 0$ since $%
\frac 1{Th^{\tau \tau }\sqrt{h}}$ never vanishes, can be used in the other
two of (\ref{vincsec.tot.1}). The equivalent system of stability condition
is: 
\begin{equation}
\left\{ 
\begin{array}{l}
\left( 
%TCIMACRO{
%\dfrac{\left( h^{\tau \sigma }\right) ^3}{h^{\sigma \sigma }\left( h^{\tau \tau }\right) ^2} }
%BeginExpansion
{\displaystyle {\left( h^{\tau \sigma }\right) ^3 \over h^{\sigma \sigma }\left( h^{\tau \tau }\right) ^2}}
%EndExpansion
-%
%TCIMACRO{\dfrac{h^{\tau \sigma }}{h^{\tau \tau }} }
%BeginExpansion
{\displaystyle {h^{\tau \sigma } \over h^{\tau \tau }}}
%EndExpansion
\right) X_\mu ^{\prime }X^{\prime \mu }+%
%TCIMACRO{\dfrac{h^{\tau \sigma }}{h^{\sigma \sigma }} }
%BeginExpansion
{\displaystyle {h^{\tau \sigma } \over h^{\sigma \sigma }}}
%EndExpansion
%TCIMACRO{\dfrac{P^\mu P_\mu }{T^2\left( h^{\tau \tau }\right) ^2h} }
%BeginExpansion
{\displaystyle {P^\mu P_\mu  \over T^2\left( h^{\tau \tau }\right) ^2h}}
%EndExpansion
\approx 0, \\ 
\\ 
\left( 1- 
%TCIMACRO{
%\dfrac{\left( h^{\tau \sigma }\right) ^2}{h^{\sigma \sigma }h^{\tau \tau }} }
%BeginExpansion
{\displaystyle {\left( h^{\tau \sigma }\right) ^2 \over h^{\sigma \sigma }h^{\tau \tau }}}
%EndExpansion
\right) X^{\prime \mu }X_\mu ^{\prime }-%
%TCIMACRO{\dfrac 1{T^2h^{\sigma \sigma }h^{\tau \tau }h} }
%BeginExpansion
{\displaystyle {1 \over T^2h^{\sigma \sigma }h^{\tau \tau }h}}
%EndExpansion
P^\mu P_\mu \approx 0, \\ 
\\ 
P_\mu X^{\prime \mu }\approx 0.
\end{array}
\right.  \label{vincsec.tot.2}
\end{equation}
We can work on the first line, getting 
\[
%TCIMACRO{\dfrac{h^{\tau \sigma }}{h^{\tau \tau }}}
%BeginExpansion
{\displaystyle {h^{\tau \sigma } \over h^{\tau \tau }}}
%EndExpansion
\left[ \left( 1-%
%TCIMACRO{
%\dfrac{\left( h^{\tau \sigma }\right) ^2}{h^{\sigma \sigma }h^{\tau \tau }}}
%BeginExpansion
{\displaystyle {\left( h^{\tau \sigma }\right) ^2 \over h^{\sigma \sigma }h^{\tau \tau }}}
%EndExpansion
\right) X_\mu ^{\prime }X^{\prime \mu }-%
%TCIMACRO{\dfrac 1{T^2h^{\sigma \sigma }h^{\tau \tau }h}}
%BeginExpansion
{\displaystyle {1 \over T^2h^{\sigma \sigma }h^{\tau \tau }h}}
%EndExpansion
P^\mu P_\mu \right] \approx 0 
\]
and, since $h^{\tau \sigma }$ must be considered free, we'll hold the
constraint: 
\[
\left( 1-%
%TCIMACRO{
%\dfrac{\left( h^{\tau \sigma }\right) ^2}{h^{\sigma \sigma }h^{\tau \tau }}}
%BeginExpansion
{\displaystyle {\left( h^{\tau \sigma }\right) ^2 \over h^{\sigma \sigma }h^{\tau \tau }}}
%EndExpansion
\right) X_\mu ^{\prime }X^{\prime \mu }-%
%TCIMACRO{\dfrac 1{T^2h^{\sigma \sigma }h^{\tau \tau }h}}
%BeginExpansion
{\displaystyle {1 \over T^2h^{\sigma \sigma }h^{\tau \tau }h}}
%EndExpansion
P^\mu P_\mu \approx 0. 
\]
The independent array of secondary constraints, allowing stability for (\ref
{vincprim}), becomes: 
\begin{equation}
\begin{array}{ccc}
\left( 1- 
%TCIMACRO{
%\dfrac{\left( h^{\tau \sigma }\right) ^2}{h^{\sigma \sigma }h^{\tau \tau }} }
%BeginExpansion
{\displaystyle {\left( h^{\tau \sigma }\right) ^2 \over h^{\sigma \sigma }h^{\tau \tau }}}
%EndExpansion
\right) X^{\prime \mu }X_\mu ^{\prime }-%
%TCIMACRO{\dfrac 1{T^2h^{\sigma \sigma }h^{\tau \tau }h} }
%BeginExpansion
{\displaystyle {1 \over T^2h^{\sigma \sigma }h^{\tau \tau }h}}
%EndExpansion
P^\mu P_\mu \approx 0, &  & P_\mu X^{\prime \mu }\approx 0.
\end{array}
\label{vincsec.tot.3}
\end{equation}
We'll use the following symbols 
\begin{equation}
\begin{array}{ccc}
\chi _1=P_\mu X^{\prime \mu }, &  & \Gamma =\left( 1- 
%TCIMACRO{
%\dfrac{\left( h^{\tau \sigma }\right) ^2}{h^{\sigma \sigma }h^{\tau \tau }} }
%BeginExpansion
{\displaystyle {\left( h^{\tau \sigma }\right) ^2 \over h^{\sigma \sigma }h^{\tau \tau }}}
%EndExpansion
\right) X^{\prime \mu }X_\mu ^{\prime }-%
%TCIMACRO{\dfrac 1{T^2h^{\sigma \sigma }h^{\tau \tau }h} }
%BeginExpansion
{\displaystyle {1 \over T^2h^{\sigma \sigma }h^{\tau \tau }h}}
%EndExpansion
P^\mu P_\mu
\end{array}
\label{v.36}
\end{equation}
for sake of simplicity. In this language, it's possible to render the
stability conditions clearer, replacing $\Gamma $ with a simpler function of 
$P$ and $X^{\prime }$: in fact, from (\ref{segn.h.1}) one writes: 
\begin{equation}
\Gamma = 
%TCIMACRO{
%\dfrac{\det \left\| h^{{\rm ab}}\right\| }{h^{\sigma \sigma }h^{\tau \tau }} }
%BeginExpansion
{\displaystyle {\det \left\| h^{{\rm ab}}\right\|  \over h^{\sigma \sigma }h^{\tau \tau }}}
%EndExpansion
\left( 
%TCIMACRO{\dfrac 1{T^2} }
%BeginExpansion
{\displaystyle {1 \over T^2}}
%EndExpansion
P^\mu P_\mu +X^{\prime \mu }X_\mu ^{\prime }\right) .  \label{w.2}
\end{equation}
Since $h^{{\rm ab}}$ is nonsingular, one recognizes that $\Gamma $ vanishes
if and only if 
\begin{equation}
\chi _2\approx 0,  \label{v.40.ter}
\end{equation}
where 
\begin{equation}
\chi _2\left( P,X^{\prime }\right) =%
%TCIMACRO{\dfrac 1{T^2} }
%BeginExpansion
{\displaystyle {1 \over T^2}}
%EndExpansion
P^\mu P_\mu +X^{\prime \mu }X_\mu ^{\prime }  \label{v.40.ter.bis}
\end{equation}
and one can use the following system of primary plus secondary constraints 
\begin{equation}
\begin{array}{ccccc}
\pi _{\tau \tau }\approx 0, & \pi _{\tau \sigma }\approx 0, & \pi _{\sigma
\sigma }\approx 0, & \chi _1\left( P,X^{\prime }\right) \approx 0, & \chi
_2\left( P,X^{\prime }\right) \approx 0.
\end{array}
\label{v.35}
\end{equation}

\subsection{Stability and Lagrange multipliers.}

Now one has to go on to work out the stability conditions for the whole set
of constraints, checking their consistency with the motion generated by the
canonical Hamiltonian linear density 
\begin{equation}
{\cal H}_{{\rm C}}=-%
%TCIMACRO{\dfrac{f\left[ h\right] }{2T} }
%BeginExpansion
{\displaystyle {f\left[ h\right]  \over 2T}}
%EndExpansion
P^\mu P_\mu -%
%TCIMACRO{\dfrac T2 }
%BeginExpansion
{\displaystyle {T \over 2}}
%EndExpansion
f\left[ h\right] X^{\prime \mu }X_\mu ^{\prime }-%
%TCIMACRO{\dfrac{h^{\tau \sigma }}{h^{\tau \tau }} }
%BeginExpansion
{\displaystyle {h^{\tau \sigma } \over h^{\tau \tau }}}
%EndExpansion
P^\mu X_\mu ^{\prime }+\lambda ^{\tau \tau }\pi _{\tau \tau }+\lambda ^{\tau
\sigma }\pi _{\tau \sigma }+\lambda ^{\sigma \sigma }\pi _{\sigma \sigma },
\label{densham.ex.1}
\end{equation}
obtained by adding to the ''constraintless'' Hamiltonian (\ref{densham.can})
a linear combination of constraints with the Lagrange multipliers. In (\ref
{densham.ex.1}) we put: 
\begin{equation}
f\left[ h\right] = 
%TCIMACRO{
%\dfrac{\sqrt{\left( h^{\tau \sigma }\right) ^2-h^{\tau \tau }h^{\sigma \sigma }}}{h^{\tau \tau }} }
%BeginExpansion
{\displaystyle {\sqrt{\left( h^{\tau \sigma }\right) ^2-h^{\tau \tau }h^{\sigma \sigma }} \over h^{\tau \tau }}}
%EndExpansion
\label{v.44}
\end{equation}
for simplicity.

In order to look for stability conditions of constraints, it's better to
express even the ''constraintless'' part in terms of the very constraints.
It's easy to recognize: 
\[
{\cal H}_0=-%
%TCIMACRO{\dfrac{h^{\tau \sigma }}{h^{\tau \tau }}}
%BeginExpansion
{\displaystyle {h^{\tau \sigma } \over h^{\tau \tau }}}
%EndExpansion
\chi _1-%
%TCIMACRO{\dfrac T2}
%BeginExpansion
{\displaystyle {T \over 2}}
%EndExpansion
f\left[ h\right] \chi _2. 
\]
Thus the linear density for the canonical Hamiltonian is: 
\begin{equation}
{\cal H}_{{\rm C}}=\lambda ^{\tau \tau }\pi _{\tau \tau }+\lambda ^{\tau
\sigma }\pi _{\tau \sigma }+\lambda ^{\sigma \sigma }\pi _{\sigma \sigma }-%
%TCIMACRO{\dfrac{h^{\tau \sigma }}{h^{\tau \tau }} }
%BeginExpansion
{\displaystyle {h^{\tau \sigma } \over h^{\tau \tau }}}
%EndExpansion
\chi _1-%
%TCIMACRO{\dfrac T2 }
%BeginExpansion
{\displaystyle {T \over 2}}
%EndExpansion
f\left[ h\right] \chi _2.  \label{densham.ex.2}
\end{equation}
{\it The scalar canonical Hamiltonian reduces to a linear combination of
constraints.}

When (\ref{v.35}) are fulfilled, one has: 
\begin{equation}
{\cal H}_{{\rm C}}\approx 0.  \label{v.39}
\end{equation}
This is exactly what happens in the relativistic free classical particle,
where $H_{{\rm C}}=\lambda \left( p^\mu p_\mu -m^2\right) \approx 0$.

With the Hamiltonian 
\begin{equation}
H_{{\rm C}}=%
%TCIMACRO{\dint }
%BeginExpansion
\displaystyle \int 
%EndExpansion
\limits_0^\pi d\sigma \left( \lambda ^{\tau \tau }\pi _{\tau \tau }+\lambda
^{\tau \sigma }\pi _{\tau \sigma }+\lambda ^{\sigma \sigma }\pi _{\sigma
\sigma }-%
%TCIMACRO{\dfrac{h^{\tau \sigma }}{h^{\tau \tau }} }
%BeginExpansion
{\displaystyle {h^{\tau \sigma } \over h^{\tau \tau }}}
%EndExpansion
\chi _1-%
%TCIMACRO{\dfrac T2 }
%BeginExpansion
{\displaystyle {T \over 2}}
%EndExpansion
f\left[ h\right] \chi _2\right)  \label{ham.tot.1}
\end{equation}
we're ready to work out stability conditions using: 
\begin{equation}
\begin{array}{ccccc}
\left\{ \pi _{{\rm ab}},H_{{\rm C}}\right\} \approx 0, &  & \left\{ \chi
_1,H_{{\rm C}}\right\} \approx 0, &  & \left\{ \chi _2,H_{{\rm C}}\right\}
\approx 0.
\end{array}
\label{v.40}
\end{equation}
First of all, it's necessary to produce the Poisson bracket algebra of the
constraints, where (\ref{v.19}) are supposed to be fulfilled {\it a priori}.

The formal definition of equal time Poisson bracket between to quantities $%
{\cal A}\left[ X,P,h,\pi \right] $ and ${\cal B}\left[ X,P,h,\pi \right] $
is: 
\begin{equation}
\begin{array}{l}
\left\{ {\cal A}\left( \tau \right) ,{\cal B}\left( \tau \right) \right\} =
\\ 
\\ 
=%
%TCIMACRO{\dint }
%BeginExpansion
\displaystyle \int 
%EndExpansion
\limits_0^\pi d\sigma \left[ 
%TCIMACRO{
%\dfrac{\delta {\cal A}\left( \tau \right) }{\delta X^\mu \left( \sigma ,\tau \right) } }
%BeginExpansion
{\displaystyle {\delta {\cal A}\left( \tau \right)  \over \delta X^\mu \left( \sigma ,\tau \right) }}
%EndExpansion
%TCIMACRO{
%\dfrac{\delta {\cal B}\left( \tau \right) }{\delta P_\mu \left( \sigma ,\tau \right) } }
%BeginExpansion
{\displaystyle {\delta {\cal B}\left( \tau \right)  \over \delta P_\mu \left( \sigma ,\tau \right) }}
%EndExpansion
- 
%TCIMACRO{
%\dfrac{\delta {\cal A}\left( \tau \right) }{\delta P_\mu \left( \sigma ,\tau \right) } }
%BeginExpansion
{\displaystyle {\delta {\cal A}\left( \tau \right)  \over \delta P_\mu \left( \sigma ,\tau \right) }}
%EndExpansion
%TCIMACRO{
%\dfrac{\delta {\cal B}\left( \tau \right) }{\delta X^\mu \left( \sigma ,\tau \right) } }
%BeginExpansion
{\displaystyle {\delta {\cal B}\left( \tau \right)  \over \delta X^\mu \left( \sigma ,\tau \right) }}
%EndExpansion
\right] + \\ 
\\ 
+%
%TCIMACRO{\dint }
%BeginExpansion
\displaystyle \int 
%EndExpansion
\limits_0^\pi d\sigma \left[ 
%TCIMACRO{
%\dfrac{\delta {\cal A}\left( \tau \right) }{\delta h^{{\rm ab}}\left( \sigma ,\tau \right) } }
%BeginExpansion
{\displaystyle {\delta {\cal A}\left( \tau \right)  \over \delta h^{{\rm ab}}\left( \sigma ,\tau \right) }}
%EndExpansion
%TCIMACRO{
%\dfrac{\delta {\cal B}\left( \tau \right) }{\delta \pi _{{\rm ab}}\left( \sigma ,\tau \right) } }
%BeginExpansion
{\displaystyle {\delta {\cal B}\left( \tau \right)  \over \delta \pi _{{\rm ab}}\left( \sigma ,\tau \right) }}
%EndExpansion
- 
%TCIMACRO{
%\dfrac{\delta {\cal A}\left( \tau \right) }{\delta \pi _{{\rm ab}}\left( \sigma ,\tau \right) } }
%BeginExpansion
{\displaystyle {\delta {\cal A}\left( \tau \right)  \over \delta \pi _{{\rm ab}}\left( \sigma ,\tau \right) }}
%EndExpansion
%TCIMACRO{
%\dfrac{\delta {\cal B}\left( \tau \right) }{\delta h^{{\rm ab}}\left( \sigma ,\tau \right) } }
%BeginExpansion
{\displaystyle {\delta {\cal B}\left( \tau \right)  \over \delta h^{{\rm ab}}\left( \sigma ,\tau \right) }}
%EndExpansion
\right] ,
\end{array}
\label{poisson.1}
\end{equation}
where functional derivatives are defined {\it \`a la Frechet} \cite
{materassi}. In particular one gets: 
\begin{equation}
\left\{ \pi _{{\rm ab}}\left( \sigma ,\tau \right) ,{\cal F}\right\} =-\frac{%
\delta {\cal F}}{\delta h^{{\rm ab}}\left( \sigma ,\tau \right) }
\label{poisson.2}
\end{equation}
from (\ref{poisson.1}), for any ${\cal F}\left[ X,P,h,\pi \right] $.

The first three constraints are directly postulated to be strongly in
involution with each other 
\begin{equation}
\left\{ \pi _{{\rm ab}}\left( \sigma ,\tau \right) ,\pi _{{\rm cd}}\left(
\sigma ^{\prime },\tau \right) \right\} =0.  \label{v.41}
\end{equation}
Poisson bracketing the momenta $\pi _{{\rm ab}}\ $with $\chi _k$ we simply
get zero, since $\chi _k$ are string-dependent only: 
\begin{equation}
\begin{array}{cc}
\left\{ \pi _{{\rm ab}}\left( \sigma ,\tau \right) ,\chi _1\left( \sigma
^{\prime },\tau \right) \right\} =0, & \left\{ \pi _{{\rm ab}}\left( \sigma
,\tau \right) ,\chi _2\left( \sigma ^{\prime },\tau \right) \right\} =0.
\end{array}
\label{w.3}
\end{equation}

Poisson brackets between $\chi _k$'s are evaluated by using the relation 
\begin{equation}
%TCIMACRO{\dfrac \partial {\partial \sigma ^{\prime }} }
%BeginExpansion
{\displaystyle {\partial \over \partial \sigma ^{\prime }}}
%EndExpansion
\delta \left( \sigma ^{\prime }-\sigma \right) =-%
%TCIMACRO{\dfrac \partial {\partial \sigma } }
%BeginExpansion
{\displaystyle {\partial \over \partial \sigma }}
%EndExpansion
\delta \left( \sigma -\sigma ^{\prime }\right)  \label{v.45}
\end{equation}
(that's a distributional equality, i.e. 
\[
\begin{array}{ccc}
%TCIMACRO{\dint }
%BeginExpansion
\displaystyle \int 
%EndExpansion
f\left( \sigma ^{\prime }\right) d\sigma ^{\prime }%
%TCIMACRO{\dfrac \partial {\partial \sigma ^{\prime }} }
%BeginExpansion
{\displaystyle {\partial \over \partial \sigma ^{\prime }}}
%EndExpansion
\delta \left( \sigma ^{\prime }-\sigma \right) =-%
%TCIMACRO{\dint }
%BeginExpansion
\displaystyle \int 
%EndExpansion
f\left( \sigma ^{\prime }\right) d\sigma ^{\prime }%
%TCIMACRO{\dfrac \partial {\partial \sigma } }
%BeginExpansion
{\displaystyle {\partial \over \partial \sigma }}
%EndExpansion
\delta \left( \sigma -\sigma ^{\prime }\right) & \forall & f),
\end{array}
\]
and read: 
\[
\left\{ 
\begin{array}{l}
\left\{ \chi _1\left( \sigma ,\tau \right) ,\chi _2\left( \sigma ^{\prime
},\tau \right) \right\} =2\left[ 
%TCIMACRO{\dfrac 1{T^2} }
%BeginExpansion
{\displaystyle {1 \over T^2}}
%EndExpansion
P^\mu \left( \sigma \right) P_\mu \left( \sigma ^{\prime }\right) +X_\mu
^{\prime }\left( \sigma \right) X^{\prime \mu }\left( \sigma ^{\prime
}\right) \right] 
%TCIMACRO{\dfrac \partial {\partial \sigma } }
%BeginExpansion
{\displaystyle {\partial \over \partial \sigma }}
%EndExpansion
\delta \left( \sigma -\sigma ^{\prime }\right) , \\ 
\\ 
\left\{ \chi _1\left( \sigma ,\tau \right) ,\chi _1\left( \sigma ^{\prime
},\tau \right) \right\} =\left[ X_\alpha ^{\prime }\left( \sigma ^{\prime
},\tau \right) P^\alpha \left( \sigma ,\tau \right) +X_\alpha ^{\prime
}\left( \sigma ,\tau \right) P^\alpha \left( \sigma ^{\prime },\tau \right)
\right] 
%TCIMACRO{\dfrac \partial {\partial \sigma } }
%BeginExpansion
{\displaystyle {\partial \over \partial \sigma }}
%EndExpansion
\delta \left( \sigma -\sigma ^{\prime }\right) , \\ 
\\ 
\left\{ \chi _2\left( \sigma ,\tau \right) ,\chi _2\left( \sigma ^{\prime
},\tau \right) \right\} =%
%TCIMACRO{\dfrac 4{T^2} }
%BeginExpansion
{\displaystyle {4 \over T^2}}
%EndExpansion
\left[ X_\alpha ^{\prime }\left( \sigma ^{\prime }\right) P^\alpha \left(
\sigma \right) +X_\alpha ^{\prime }\left( \sigma \right) P^\alpha \left(
\sigma ^{\prime }\right) \right] 
%TCIMACRO{\dfrac \partial {\partial \sigma } }
%BeginExpansion
{\displaystyle {\partial \over \partial \sigma }}
%EndExpansion
\delta \left( \sigma -\sigma ^{\prime }\right) .
\end{array}
\right. 
\]
They mean that those Poisson brackets are weakly zero, since from Dirac
function properties one discovers: 
\begin{equation}
F\left( \sigma ^{\prime }\right) 
%TCIMACRO{\dfrac \partial {\partial \sigma } }
%BeginExpansion
{\displaystyle {\partial \over \partial \sigma }}
%EndExpansion
\delta \left( \sigma -\sigma ^{\prime }\right) =%
%TCIMACRO{\dfrac \partial {\partial \sigma } }
%BeginExpansion
{\displaystyle {\partial \over \partial \sigma }}
%EndExpansion
\left[ F\left( \sigma \right) \delta \left( \sigma -\sigma ^{\prime }\right)
\right] -\delta \left( \sigma -\sigma ^{\prime }\right) 
%TCIMACRO{\dfrac \partial {\partial \sigma } }
%BeginExpansion
{\displaystyle {\partial \over \partial \sigma }}
%EndExpansion
F\left( \sigma \right)  \label{delta}
\end{equation}
which leads to 
\begin{equation}
\left\{ 
\begin{array}{l}
\left\{ \chi _1\left( \sigma ,\tau \right) ,\chi _2\left( \sigma ^{\prime
},\tau \right) \right\} =2%
%TCIMACRO{\dfrac \partial {\partial \sigma } }
%BeginExpansion
{\displaystyle {\partial \over \partial \sigma }}
%EndExpansion
\left[ \chi _2\left( \sigma \right) \delta \left( \sigma -\sigma ^{\prime
}\right) \right] -2\delta \left( \sigma -\sigma ^{\prime }\right) 
%TCIMACRO{\dfrac \partial {\partial \sigma } }
%BeginExpansion
{\displaystyle {\partial \over \partial \sigma }}
%EndExpansion
\chi _2\left( \sigma \right) , \\ 
\\ 
\left\{ \chi _1\left( \sigma ,\tau \right) ,\chi _1\left( \sigma ^{\prime
},\tau \right) \right\} =2%
%TCIMACRO{\dfrac \partial {\partial \sigma } }
%BeginExpansion
{\displaystyle {\partial \over \partial \sigma }}
%EndExpansion
\left[ \chi _1\left( \sigma \right) \delta \left( \sigma -\sigma ^{\prime
}\right) \right] -2\delta \left( \sigma -\sigma ^{\prime }\right) 
%TCIMACRO{\dfrac \partial {\partial \sigma } }
%BeginExpansion
{\displaystyle {\partial \over \partial \sigma }}
%EndExpansion
\chi _1\left( \sigma \right) , \\ 
\\ 
\left\{ \chi _2\left( \sigma ,\tau \right) ,\chi _2\left( \sigma ^{\prime
},\tau \right) \right\} =%
%TCIMACRO{\dfrac 8{T^2} }
%BeginExpansion
{\displaystyle {8 \over T^2}}
%EndExpansion
%TCIMACRO{\dfrac \partial {\partial \sigma } }
%BeginExpansion
{\displaystyle {\partial \over \partial \sigma }}
%EndExpansion
\left[ \chi _1\left( \sigma \right) \delta \left( \sigma -\sigma ^{\prime
}\right) \right] -%
%TCIMACRO{\dfrac 8{T^2} }
%BeginExpansion
{\displaystyle {8 \over T^2}}
%EndExpansion
\delta \left( \sigma -\sigma ^{\prime }\right) 
%TCIMACRO{\dfrac \partial {\partial \sigma } }
%BeginExpansion
{\displaystyle {\partial \over \partial \sigma }}
%EndExpansion
\chi _1\left( \sigma \right)
\end{array}
\right.  \label{v.45.bis}
\end{equation}
in our specific case. So one can recognize: 
\begin{equation}
\left\{ 
\begin{array}{l}
\begin{array}{ccc}
\left\{ \pi _{{\rm ab}}\left( \sigma ,\tau \right) ,\pi _{{\rm cd}}\left(
\sigma ^{\prime },\tau \right) \right\} =0, &  & \left\{ \pi _{{\rm ab}%
}\left( \sigma ,\tau \right) ,\chi _i\left( \sigma ^{\prime },\tau \right)
\right\} \approx 0,
\end{array}
\\ 
\\ 
\left\{ \chi _1\left( \sigma ,\tau \right) ,\chi _2\left( \sigma ^{\prime
},\tau \right) \right\} \approx 0, \\ 
\\ 
\begin{array}{ccc}
\left\{ \chi _1\left( \sigma ,\tau \right) ,\chi _1\left( \sigma ^{\prime
},\tau \right) \right\} \approx 0, &  & \left\{ \chi _2\left( \sigma ,\tau
\right) ,\chi _2\left( \sigma ^{\prime },\tau \right) \right\} \approx 0.
\end{array}
\end{array}
\right.  \label{poisson.vinc.tot}
\end{equation}

From the nature of $H_{{\rm C}}$ (that's made of constraints only) one can
consider stability conditions (\ref{v.40}) fulfilled. In fact, if we
evaluate Poisson bracket of a constraint, say $\psi _\alpha $, with the
product $F\left[ \psi \right] \psi _\beta $ (which $H_{{\rm C}}$ is made
of), we get: 
\[
\left\{ \psi _\alpha ,F\left[ \psi \right] \psi _\beta \right\} =F\left[
\psi \right] \left\{ \psi _\alpha ,\psi _\beta \right\} +\psi _\beta \left\{
\psi _\alpha ,F\left[ \psi \right] \right\} \approx 0, 
\]
as it was to be shown.

\subsection{Hamiltonian equations of motion.}

Here we work out, for sake of completeness, the Hamiltonian equations of
motion for sheet variables as well as string variables, and this could be
done simply by using the {\it canonical Hamiltonian} linear density 
\begin{equation}
{\cal H}_{{\rm C}}=\lambda ^{{\rm ab}}\pi _{{\rm ba}}-%
%TCIMACRO{\dfrac{h^{\tau \sigma }}{h^{\tau \tau }} }
%BeginExpansion
{\displaystyle {h^{\tau \sigma } \over h^{\tau \tau }}}
%EndExpansion
\chi _1-%
%TCIMACRO{\dfrac T2 }
%BeginExpansion
{\displaystyle {T \over 2}}
%EndExpansion
f\left[ h\right] \chi _2,  \label{v.72}
\end{equation}
which includes primary constraints only. Secondary constraints can be added
as well (as Hennaux and Teitelboim suggest in \cite{Teitelboim}), getting
the {\it extended Hamiltonian} 
\begin{equation}
{\cal H}_{{\rm E}}=\lambda ^{{\rm ab}}\pi _{{\rm ba}}+\left( \lambda _1-%
%TCIMACRO{\dfrac{h^{\tau \sigma }}{h^{\tau \tau }} }
%BeginExpansion
{\displaystyle {h^{\tau \sigma } \over h^{\tau \tau }}}
%EndExpansion
\right) \chi _1+\left( \lambda _2-%
%TCIMACRO{\dfrac T2 }
%BeginExpansion
{\displaystyle {T \over 2}}
%EndExpansion
f\left[ h\right] \right) \chi _2:  \label{v.73}
\end{equation}
generating the motion with this ${\cal H}_{{\rm E}}$ it's more evident how
the presence of secondary constraints renders ambiguous the motion of $P$
and $X$ within the coordinate-momenta manifold.

In terms of Poisson brackets, Hamiltonian equations of motion read: 
\begin{equation}
\left\{ 
\begin{array}{l}
\begin{array}{ccc}
\dot h^{{\rm ab}}=\left\{ h^{{\rm ab}},H\right\} , &  & \dot \pi _{{\rm ab}%
}=\left\{ \pi _{{\rm ab}},H\right\} ,
\end{array}
\\ 
\\ 
\begin{array}{ccc}
\dot X^\mu =\left\{ X^\mu ,H\right\} , &  & \dot P^\mu =\left\{ P^\mu
,H\right\} ,
\end{array}
\end{array}
\right.  \label{v.74}
\end{equation}
where $H$ is obtained by integrating (\ref{v.72}) along the string, 
\[
\left\{ 
\begin{array}{l}
\begin{array}{ccc}
\dot h^{{\rm ab}}=\left\{ h^{{\rm ab}},H_{{\rm C}}\right\} , &  & \dot \pi _{%
{\rm ab}}=\left\{ \pi _{{\rm ab}},H_{{\rm C}}\right\} ,
\end{array}
\\ 
\\ 
\begin{array}{ccc}
\dot X^\mu =\left\{ X^\mu ,H_{{\rm C}}\right\} , &  & \dot P^\mu =\left\{
P^\mu ,H_{{\rm C}}\right\} ,
\end{array}
\end{array}
\right. 
\]
or (\ref{v.73}) 
\[
\left\{ 
\begin{array}{l}
\begin{array}{ccc}
\dot h^{{\rm ab}}=\left\{ h^{{\rm ab}},H_{{\rm E}}\right\} , &  & \dot \pi _{%
{\rm ab}}=\left\{ \pi _{{\rm ab}},H_{{\rm E}}\right\} ,
\end{array}
\\ 
\\ 
\begin{array}{ccc}
\dot X^\mu =\left\{ X^\mu ,H_{{\rm E}}\right\} , &  & \dot P^\mu =\left\{
P^\mu ,H_{{\rm E}}\right\} .
\end{array}
\end{array}
\right. 
\]
We'll use directly the {\it extended Hamiltonian} 
\begin{equation}
H_{{\rm E}}=%
%TCIMACRO{\dint }
%BeginExpansion
\displaystyle \int 
%EndExpansion
d\sigma \left[ \lambda ^{{\rm ab}}\pi _{{\rm ba}}+\left( \lambda _1-%
%TCIMACRO{\dfrac{h^{\tau \sigma }}{h^{\tau \tau }} }
%BeginExpansion
{\displaystyle {h^{\tau \sigma } \over h^{\tau \tau }}}
%EndExpansion
\right) \chi _1+\left( \lambda _2-%
%TCIMACRO{\dfrac T2 }
%BeginExpansion
{\displaystyle {T \over 2}}
%EndExpansion
f\left[ h\right] \right) \chi _2\right]  \label{v.75}
\end{equation}
and we'll consider the rules (\ref{v.19}) and (\ref{v.66}).

Sheet variables obey the following equations 
\[
\dot h^{{\rm ab}}\left( \sigma ,\tau \right) =%
%TCIMACRO{\dfrac 12}
%BeginExpansion
{\displaystyle {1 \over 2}}
%EndExpansion
\left[ \lambda ^{{\rm ab}}\left( \sigma ,\tau \right) +\lambda ^{{\rm ba}%
}\left( \sigma ,\tau \right) \right] , 
\]
but the symmetry $\pi _{{\rm ba}}=\pi _{{\rm ab}}$ allows only the symmetric
part of $\lambda ^{{\rm ab}}$ take part to the play, so that: 
\begin{equation}
\dot h^{{\rm ab}}\left( \sigma ,\tau \right) =\lambda ^{{\rm ab}}\left(
\sigma ,\tau \right) .  \label{v.76}
\end{equation}
The former expresses physical emptiness of $h^{{\rm ab}}$: it can be used to
invert equation (\ref{momento.h}), and gives us an idea of arbitrariness of $%
h^{{\rm ab}}$ motion.

Hamilton's equations for sheet momenta are: 
\begin{equation}
\dot \pi _{{\rm ab}}\left( \sigma ,\tau \right) \approx 0,  \label{v.77}
\end{equation}
and represent the stability of $\pi _{{\rm ab}}$ as constraints.

Let's deal with string variables; the Lagrangean coordinates obey to: 
\begin{equation}
\dot X_\mu \left( \sigma ,\tau \right) =\left( \lambda _1\left( \sigma ,\tau
\right) - 
%TCIMACRO{
%\dfrac{h^{\tau \sigma }\left( \sigma ,\tau \right) }{h^{\tau \tau }\left( \sigma ,\tau \right) } }
%BeginExpansion
{\displaystyle {h^{\tau \sigma }\left( \sigma ,\tau \right)  \over h^{\tau \tau }\left( \sigma ,\tau \right) }}
%EndExpansion
\right) X_\mu ^{\prime }\left( \sigma ,\tau \right) +\left( \frac{2\lambda
_2\left( \sigma ,\tau \right) }{T^2}-%
%TCIMACRO{\dfrac{f\left[ h\left( \sigma ,\tau \right) \right] }T }
%BeginExpansion
{\displaystyle {f\left[ h\left( \sigma ,\tau \right) \right]  \over T}}
%EndExpansion
\right) P_\mu \left( \sigma ,\tau \right) .  \label{v.78}
\end{equation}
This represents the inversion formula for (\ref{momento.x}), in which
Lagrange multipliers $\lambda _1\left( \sigma ,\tau \right) $ and $\lambda
_2\left( \sigma ,\tau \right) $ appear: equation (\ref{v.78}) does coincide
with (\ref{v.23}) when multipliers are chosen to vanish.

The real physically meaningful equation of motion for the string variables
constructed with extended Hamiltonian is anyway that of $P$: 
\begin{equation}
\begin{array}{l}
\dot P_\mu \left( \sigma ,\tau \right) = \\ 
\\ 
=\left[ \lambda _1^{\prime }\left( \sigma ,\tau \right) -%
%TCIMACRO{\dfrac \partial {\partial \sigma } }
%BeginExpansion
{\displaystyle {\partial \over \partial \sigma }}
%EndExpansion
\left( 
%TCIMACRO{
%\dfrac{h^{\tau \sigma }\left( \sigma ,\tau \right) }{h^{\tau \tau }\left( \sigma ,\tau \right) } }
%BeginExpansion
{\displaystyle {h^{\tau \sigma }\left( \sigma ,\tau \right)  \over h^{\tau \tau }\left( \sigma ,\tau \right) }}
%EndExpansion
\right) \right] P_\mu \left( \sigma ,\tau \right) +\left( \lambda _1\left(
\sigma ,\tau \right) - 
%TCIMACRO{
%\dfrac{h^{\tau \sigma }\left( \sigma ,\tau \right) }{h^{\tau \tau }\left( \sigma ,\tau \right) } }
%BeginExpansion
{\displaystyle {h^{\tau \sigma }\left( \sigma ,\tau \right)  \over h^{\tau \tau }\left( \sigma ,\tau \right) }}
%EndExpansion
\right) P_\mu ^{\prime }\left( \sigma ,\tau \right) + \\ 
\\ 
+\left( 2\lambda _2^{\prime }\left( \sigma ,\tau \right) -Tf^{\prime }\left[
h\left( \sigma ,\tau \right) \right] \right) X_\mu ^{\prime }\left( \sigma
,\tau \right) +\left( 2\lambda _2\left( \sigma ,\tau \right) -Tf\left[
h\left( \sigma ,\tau \right) \right] \right) X_\mu ^{\prime \prime }\left(
\sigma ,\tau \right) .
\end{array}
\label{v.79}
\end{equation}
Lagrange multipliers appear here too, with their degree of arbitrariness.

\section{Symmetries and constraints.}

Let's now analyze the meaning of constraints shown by Polyakov's string,
treating them as canonical generators of gauge transformations.

\subsection{Gauge transformations for sheet variables.}

The physical system with action 
\begin{equation}
S=-\frac T2%
%TCIMACRO{\dint }
%BeginExpansion
\displaystyle \int 
%EndExpansion
d\tau 
%TCIMACRO{\dint }
%BeginExpansion
\displaystyle \int 
%EndExpansion
d\sigma \sqrt{h}h^{{\rm ab}}\partial _{{\rm a}}X^\mu \partial _{{\rm b}}X_\mu
\label{azione.pol.1}
\end{equation}
shows five first class constraints 
\[
\begin{array}{ccc}
\left\{ 
\begin{array}{l}
\pi _{\tau \tau }\approx 0, \\ 
\\ 
\pi _{\tau \sigma }\approx 0, \\ 
\\ 
\pi _{\sigma \sigma }\approx 0,
\end{array}
\right. &  & 
\begin{array}{ccc}
P^\mu X_\mu ^{\prime }\approx 0, &  & 
%TCIMACRO{\dfrac{P^\alpha P_\alpha }{T^2} }
%BeginExpansion
{\displaystyle {P^\alpha P_\alpha  \over T^2}}
%EndExpansion
+X^{\prime \alpha }X_\alpha ^{\prime }\approx 0,
\end{array}
\end{array}
\]
three primary and two secondary.

Since the canonical momenta of variables $h^{{\rm ab}}$ representing the
worldsheet metric are constraints, these variables are functions of $\sigma $
and $\tau $ which can be arbitrarily changed at each instant of the motion,
without changing the physical state of the system. In particular, we've
already shown in (\ref{v.76}) that Legendre transformations from $\dot h^{%
{\rm ab}}$ to $\pi _{{\rm ab}}$ must be inverted inserting arbitrary scalar
functions of $\tau $.

It's possible to use this arbitrariness affecting $h^{{\rm ab}}$ by
operating some gauge fixing in order to make simpler the string description:
we'll now perform this gauge-fixing.

This operation needs some mathematical conditions:\ gauge fixing is the
position of some conditions\footnote{%
...which must fulfill \cite{Teitelboim}:
\par
\begin{itemize}
\item  {\it Accessibility}: from the original form of $h^{{\rm ab}}$ it must
be possible to reach a new form $h_0^{{\rm ab}}$ identically satisfying (\ref
{v.80}). This must be done with a sequence of transformations generated by
suitable second class constraints, $\pi _{{\rm ab}}$ in our present case.
\par
\item  {\it Completeness}: relationships (\ref{v.80}) must completely fix
the gauge, i.e. the form of $h_0^{{\rm ab}}$ we've given the sheet variables
must be uninvariant under those transformations of the gauge we wanted to
fix; in particular this condition is formulated as 
\begin{equation}
\left\{ C_n\left[ h^{{\rm ab}}\right] ,\pi _{{\rm ef}}\right\} \neq 0,
\label{v.81}
\end{equation}
in terms of Poisson brackets.
\end{itemize}
} 
\begin{equation}
C_n\left[ h^{{\rm ab}}\right] =0.  \label{v.80}
\end{equation}

Let's start with a particular metric configuration 
\begin{equation}
h^{{\rm ab}}=\tilde h^{{\rm ab}}  \label{v.82}
\end{equation}
being $\tilde h^{{\rm ab}}$ assigned, and let's look for the $\pi _{{\rm ef}%
} $-generated transformation leading to the wanted form $h_0^{{\rm ab}}$.

Let's define the functional derivative differential operator $\Pi _{{\rm ef}%
} $, acting on quantities depending from sheet variables, such that: 
\begin{equation}
\Pi _{{\rm ef}}\left( \sigma \right) {\cal F}\left[ h\right] =\left\{ {\cal F%
}\left[ h\right] ,\pi _{{\rm ef}}\left( \sigma \right) \right\} ,
\label{v.83}
\end{equation}
and let's appreciate: 
\[
\Pi _{{\rm ef}}\left( \sigma \right) {\cal F}\left[ h\right] =\frac{\delta 
{\cal F}\left[ h\right] }{\delta h^{{\rm ef}}\left( \sigma \right) }. 
\]
We can thus use the operator identification: 
\begin{equation}
\Pi _{{\rm ef}}\left( \sigma \right) =\frac \delta {\delta h^{{\rm ef}%
}\left( \sigma \right) }.  \label{v.84}
\end{equation}

Any infinitesimal element of $\pi _{{\rm ef}}$-algebra 
\[
%TCIMACRO{\dint }
%BeginExpansion
\displaystyle \int 
%EndExpansion
\epsilon ^{{\rm ef}}\left( \sigma \right) \pi _{{\rm ef}}\left( \sigma
\right) d\sigma 
\]
will cause a transformation 
\[
{\cal F}\left[ h\right] \rightarrow {\cal F}\left[ h\right] +\delta {\cal F}%
\left[ h\right] 
\]
on any functional ${\cal F}\left[ h\right] $, such that: 
\begin{equation}
\delta {\cal F}\left[ h\right] =%
%TCIMACRO{\dint }
%BeginExpansion
\displaystyle \int 
%EndExpansion
\epsilon ^{{\rm ef}}\left( \sigma \right) \left\{ {\cal F}\left[ h\right]
,\pi _{{\rm ef}}\left( \sigma \right) \right\} d\sigma .  \label{v.86}
\end{equation}
This will be applied in the form (\ref{v.84}), getting: 
\begin{equation}
\delta {\cal F}\left[ h\right] =%
%TCIMACRO{\dint }
%BeginExpansion
\displaystyle \int 
%EndExpansion
d\sigma \epsilon ^{{\rm ef}}\left( \sigma \right) \frac{\delta {\cal F}%
\left[ h\right] }{\delta h^{{\rm ef}}\left( \sigma \right) }.
\label{v.86.bis}
\end{equation}
Infinitesimal variation on the metric is then: 
\begin{equation}
\delta h^{{\rm ab}}\left( \sigma \right) =\epsilon ^{{\rm ab}}\left( \sigma
\right) .  \label{v.87}
\end{equation}

Equations (\ref{v.86}) and (\ref{v.86.bis}) can be extended to finite forms,
simply by exponentiating the infinitesimal version: 
\begin{equation}
{\cal F}^{\prime }\left[ h\right] ={\cal F}\left[ h\right] +%
%TCIMACRO{\dsum }
%BeginExpansion
\mathop{\displaystyle \sum }
%EndExpansion
\limits_{n=1}^{+\infty }\frac 1{n!}\left( 
%TCIMACRO{\dint }
%BeginExpansion
\displaystyle \int 
%EndExpansion
d\sigma \epsilon ^{{\rm ef}}\left( \sigma \right) \frac \delta {\delta h^{%
{\rm ef}}\left( \sigma \right) }\right) ^n{\cal F}\left[ h\right] ,
\label{v.88}
\end{equation}
i.e.: 
\begin{equation}
{\cal F}^{\prime }=\left[ \exp \left( 
%TCIMACRO{\dint }
%BeginExpansion
\displaystyle \int 
%EndExpansion
d\sigma \epsilon ^{{\rm ef}}\left( \sigma \right) \frac \delta {\delta h^{%
{\rm ef}}\left( \sigma \right) }\right) \right] {\cal F}.  \label{v.89}
\end{equation}
These canonical generators can be very easily exponentiated, since their
bracket algebra is trivial. The fact that their algebra is abelian will be a
key point all over our discussion\footnote{%
Of course one should be careful about the convergence of the integrals
involved; we will not be so careful here, this paper is simply intended to
be a solved exercise!}.

When we work with functions from ${\Bbb R}$ into ${\Bbb R}$ we have 
\begin{equation}
\left( D_af\right) \left( y\right) =f\left( y+a\right) .  \label{v.92}
\end{equation}
where: 
\begin{equation}
D_a=%
%TCIMACRO{\dsum }
%BeginExpansion
\mathop{\displaystyle \sum }
%EndExpansion
\limits_{n=0}^{+\infty }\frac 1{n!}a^n\partial _x^n,  \label{v.91}
\end{equation}
and $a$ is a fixed real number. The following mathematical object 
\begin{equation}
{\cal D}_\epsilon =\exp \left( 
%TCIMACRO{\dint }
%BeginExpansion
\displaystyle \int 
%EndExpansion
d\sigma \epsilon ^{{\rm ef}}\left( \sigma \right) \frac \delta {\delta h^{%
{\rm ef}}\left( \sigma \right) }\right)  \label{v.93}
\end{equation}
is a thing much more sophisticated than $D_a$, here we make functional
derivatives, and have to deal with an infinite number of degrees of
freedom... Anyway, we still exponentiate abelian operators $\frac \delta {%
\delta h^{{\rm ef}}\left( \sigma \right) }$, with coefficients $\epsilon ^{%
{\rm ef}}\left( \sigma \right) $ which are functionally constant with
respect to the variables $h^{{\rm ef}}\left( \sigma \right) $. We arrive to
the {\it na\"\i ve} conclusion: 
\begin{equation}
\begin{array}{ccc}
{\cal F}^{\prime }\left[ h\right] ={\cal D}_\epsilon {\cal F}\left[ h\right]
& \Rightarrow & {\cal F}^{\prime }\left[ h\right] ={\cal F}\left[ h+\epsilon
\right] .
\end{array}
\label{v.94}
\end{equation}

Metric sheet tensor changes from $h$ into $h+\epsilon $, so that if we want
to go from $\tilde h^{{\rm ab}}$ to $h_0^{{\rm ab}}$ fulfilling (\ref{v.80})
it will be possible to do it by that ${\cal D}_\epsilon $ with 
\begin{equation}
\epsilon ^{{\rm ef}}\left( \sigma \right) =h_0^{{\rm ef}}\left( \sigma
\right) -\tilde h^{{\rm ef}}\left( \sigma \right) .  \label{v.95}
\end{equation}

\subsection{Some interesting functionals.}

So our system is described by a linear canonical Hamiltonian density 
\begin{equation}
{\cal H}_{{\rm C}}=\lambda ^{\tau \tau }\pi _{\tau \tau }+\lambda ^{\tau
\sigma }\pi _{\tau \sigma }+\lambda ^{\sigma \sigma }\pi _{\sigma \sigma }-%
%TCIMACRO{\dfrac{h^{\tau \sigma }}{h^{\tau \tau }} }
%BeginExpansion
{\displaystyle {h^{\tau \sigma } \over h^{\tau \tau }}}
%EndExpansion
\chi _1-%
%TCIMACRO{\dfrac T2 }
%BeginExpansion
{\displaystyle {T \over 2}}
%EndExpansion
f\left[ h\right] \chi _2,  \label{v.56}
\end{equation}
where $f\left[ h\right] $ is defined in (\ref{v.44}). There are five first
class constraints, three involving only {\it sheet variables} 
\begin{equation}
\begin{array}{ccc}
\pi _{\tau \tau }\approx 0, & \pi _{\tau \sigma }\approx 0, & \pi _{\sigma
\sigma }\approx 0,
\end{array}
\label{v.57}
\end{equation}
and two involving only {\it string variables}: 
\begin{equation}
\begin{array}{ccc}
P^\mu X_\mu ^{\prime }\approx 0, &  & 
%TCIMACRO{\dfrac{P^\alpha P_\alpha }{T^2} }
%BeginExpansion
{\displaystyle {P^\alpha P_\alpha  \over T^2}}
%EndExpansion
+X^{\prime \alpha }X_\alpha ^{\prime }\approx 0.
\end{array}
\label{v.58}
\end{equation}

From singular Lagrangean system theory \cite{Teitelboim}, one knows that 
{\it first class constraints are gauge transformation generators}, which
change in a $\tau $-dependent way the formal motion through the phase space,
without changing the physical state of the system. Here we'll underline
which transformation is generated by each constraint.

The action (\ref{azione.pol}) shows lots of symmetries. It is invariant
under the worldsheet reparametrizations 
\begin{equation}
\begin{array}{ccc}
\tilde \sigma =\tilde \sigma \left( \sigma ,\tau \right) , &  & \tilde \tau =%
\tilde \tau \left( \sigma ,\tau \right)
\end{array}
\label{v.58.bis}
\end{equation}
(diffeomophic maps from ${\Bbb V}_2$ into ${\Bbb V}_2$), as well as under
Weyl transformations 
\begin{equation}
\tilde h^{{\rm ab}}\left( \sigma ,\tau \right) =\Lambda \left( \sigma ,\tau
\right) h^{{\rm ab}}\left( \sigma ,\tau \right) .  \label{Weyl}
\end{equation}

The term $h^{{\rm ab}}\partial _{{\rm a}}X^\mu \partial _{{\rm b}}X^\nu $ is
diff-invariant since the sheet-tensor indices are all correctly saturated;
moreover the measure 
\begin{equation}
d^2{\Bbb V}=d\tau d\sigma \sqrt{h}  \label{misura inviariante}
\end{equation}
is notoriously a diff-invariant one, so $S$ does be.

We have 
\begin{equation}
\sqrt{h^{\prime }}\left( h^{\prime }\right) ^{{\rm ab}}=\sqrt{h}h^{{\rm ab}}
\label{v.60}
\end{equation}
changing the metric as suggested in (\ref{Weyl}), while $-\frac T2\eta _{\mu
\nu }d\tau d\sigma \partial _{{\rm a}}X^\mu \partial _{{\rm b}}X^\nu $ is
really unaffected by Weyl transformation (\ref{Weyl}), so that $S$ in
invariant under local rescaling too.

We can try to understand these symmetries in terms of transformations
generated by the first class constraints (\ref{v.57}) and (\ref{v.58}). We
have to be particularly careful with diffeomorphisms, because the
constraints act only on the fields $h$, $\pi $, $X$ and $P$, not on the $%
{\Bbb V}_2$-coordinates directly; we can still map reparametrizations (\ref
{v.58.bis}) into the corresponding transformations which affect the fields
as a consequence of those coordinate changes. For example, letting $\sigma ^{%
{\rm a}}$ be any ''old'' worldsheet variable and $\sigma ^{\prime {\rm a}}$
any ''new'' one, we can still recognize: 
\begin{equation}
\begin{array}{cccc}
h^{\prime {\rm ab}}= 
%TCIMACRO{
%\dfrac{\partial \sigma ^{\prime {\rm a}}}{\partial \sigma ^{{\rm c}}} }
%BeginExpansion
{\displaystyle {\partial \sigma ^{\prime {\rm a}} \over \partial \sigma ^{{\rm c}}}}
%EndExpansion
%TCIMACRO{
%\dfrac{\partial \sigma ^{\prime {\rm b}}}{\partial \sigma ^{{\rm d}}} }
%BeginExpansion
{\displaystyle {\partial \sigma ^{\prime {\rm b}} \over \partial \sigma ^{{\rm d}}}}
%EndExpansion
h^{{\rm cd}}, & \pi _{{\rm cd}}^{\prime }= 
%TCIMACRO{
%\dfrac{\partial \sigma ^{{\rm a}}}{\partial \sigma ^{\prime {\rm c}}} }
%BeginExpansion
{\displaystyle {\partial \sigma ^{{\rm a}} \over \partial \sigma ^{\prime {\rm c}}}}
%EndExpansion
%TCIMACRO{
%\dfrac{\partial \sigma ^{{\rm b}}}{\partial \sigma ^{\prime {\rm d}}} }
%BeginExpansion
{\displaystyle {\partial \sigma ^{{\rm b}} \over \partial \sigma ^{\prime {\rm d}}}}
%EndExpansion
\pi _{{\rm ab}}, & X^{\prime }=X, & P^{\prime }=P.
\end{array}
\label{v.60.bis}
\end{equation}

Let's deal with {\it sheet constraints} (\ref{v.57}); from: 
\begin{equation}
\left\{ \pi _{{\rm ab}}\left( \sigma ,\tau \right) ,{\cal F}\right\} =-\frac{%
\delta {\cal F}}{\delta h^{{\rm ab}}\left( \sigma ,\tau \right) }
\label{v.59}
\end{equation}
it's easy to regard $\pi _{{\rm ab}}$ as the canonical generators of
''translations along $h^{{\rm ab}}\left( \sigma ,\tau \right) $''. Let's use
these sheet constraints to realize Weyl rescaling generators.

One has to get 
\begin{equation}
\tilde h^{{\rm ab}}\left( \sigma ,\tau \right) =e^{\Lambda \left( \sigma
,\tau \right) }h^{{\rm ab}}\left( \sigma ,\tau \right)  \label{weyl.1}
\end{equation}
on sheet variables, while nothing has to happen to the string variables.

Let's consider an ${\Bbb O}\left( \Lambda \right) $ version of (\ref{weyl.1}%
) 
\[
\tilde h^{{\rm ab}}\left( \sigma ,\tau \right) =\left[ 1+\Lambda \left(
\sigma ,\tau \right) +...\right] h^{{\rm ab}}\left( \sigma ,\tau \right) , 
\]
so that the sheet metric tensor changes as: 
\begin{equation}
\delta h^{{\rm ab}}=\Lambda h^{{\rm ab}}.  \label{weyl.2}
\end{equation}
The infinitesimal generator of these transformations is a functional ${\cal W%
}_\Lambda $ such that: 
\begin{equation}
\delta h^{{\rm ab}}=\left\{ h^{{\rm ab}},{\cal W}_\Lambda \right\} .
\label{weyl.3}
\end{equation}
Let's try the function 
\begin{equation}
w_\Lambda \left( \sigma ,\tau \right) =\Lambda \left( \sigma ,\tau \right)
h^{{\rm ab}}\left( \sigma ,\tau \right) \pi _{{\rm ba}}\left( \sigma ,\tau
\right)  \label{v.61}
\end{equation}
with $\Lambda \in C^\infty \left( {\Bbb V}_2,{\Bbb R}\right) $; one has: 
\begin{equation}
\left\{ h^{{\rm ef}}\left( \sigma ,\tau \right) ,w_\Lambda \left( \sigma
^{\prime },\tau \right) \right\} =\Lambda \left( \sigma ^{\prime },\tau
\right) h^{{\rm ef}}\left( \sigma ,\tau \right) \delta \left( \sigma -\sigma
^{\prime }\right) .  \label{v.62}
\end{equation}
We get the right functional if we define 
\begin{equation}
{\cal W}_\Lambda \left[ h,\pi \right] =%
%TCIMACRO{\dint }
%BeginExpansion
\displaystyle \int 
%EndExpansion
\limits_0^\pi \Lambda \left( \sigma ,\tau \right) h^{{\rm ab}}\left( \sigma
,\tau \right) \pi _{{\rm ba}}\left( \sigma ,\tau \right) d\sigma
\label{v.63}
\end{equation}
and obtain: 
\begin{equation}
\left\{ h^{{\rm ef}}\left( \sigma ,\tau \right) ,{\cal W}_\Lambda \right\}
=\Lambda \left( \sigma ,\tau \right) h^{{\rm ef}}\left( \sigma ,\tau \right)
,  \label{v.64}
\end{equation}
which authorizes us to state:

\begin{itemize}
\item  ${\cal W}_\Lambda \left[ h,\pi \right] ${\it \ is the canonical
generator of Weyl rescalings}.
\end{itemize}

It's interesting to find that Weyl invariance of the classical theory is
related to the energy-stress sheet tensor of the string. Let us define that
tensor as in \cite{Carroll} 
\begin{equation}
T_{{\rm ab}}=-\frac 2{T\sqrt{h}}\frac{\partial {\cal L}}{\partial h^{{\rm ab}%
}};  \label{traccia.1}
\end{equation}
than let us assume the primary constraints 
\begin{equation}
\pi _{{\rm ab}}\approx 0:  \label{traccia.2.bis}
\end{equation}
the derivative 
\[
\dot w_\Lambda =\dot \Lambda h^{{\rm ab}}\pi _{{\rm ba}}+\Lambda \dot h^{%
{\rm ab}}\pi _{{\rm ba}}+\Lambda h^{{\rm ab}}\dot \pi _{{\rm ba}}, 
\]
thus becomes: 
\begin{equation}
\dot w_\Lambda \approx \Lambda h^{{\rm ab}}\dot \pi _{{\rm ba}}.
\label{traccia.3.bis}
\end{equation}
The condition in order for $\dot w_\Lambda \approx 0$ to be fulfilled (that
is: in order for the Weyl generating charge ${\cal W}_\Lambda $ to be
conserved, and so for the theory to be consistently Weyl-invariant) is 
\[
\dot \pi _{{\rm ba}}=0, 
\]
which becomes 
\[
\frac{\partial {\cal L}}{\partial h^{{\rm ba}}}=0 
\]
due to the Lagrange equations (\ref{v.2}) and Lagrangean singularity (\ref
{v.20}). From (\ref{traccia.1}) one gets: 
\begin{equation}
\begin{array}{ccc}
\dot w_\Lambda \left( \sigma ,\tau \right) \approx -%
%TCIMACRO{\dfrac T2 }
%BeginExpansion
{\displaystyle {T \over 2}}
%EndExpansion
\sqrt{h\left( \sigma ,\tau \right) }\Lambda \left( \sigma ,\tau \right) h^{%
{\rm ab}}\left( \sigma ,\tau \right) T_{{\rm ba}}\left( \sigma ,\tau \right)
& \forall & \Lambda \in C^\infty \left( {\Bbb V}_2,{\Bbb R}\right) ,
\end{array}
\label{traccia.4}
\end{equation}
which allows the adfirmation: {\it the vanishing of stress-energy tensor
trace is the condition for the local rescaling Weyl transformations to be
symmetries of the theory, because it's the condition for }${\cal W}_\Lambda $%
{\it \ to be constant.}

Let's now deal with string variable constraints, those $\chi _k$'s defined
as follows: 
\begin{equation}
\begin{array}{cc}
\chi _1=P^\mu X_\mu ^{\prime }, & \chi _2=%
%TCIMACRO{\dfrac{P^\alpha P_\alpha }{T^2} }
%BeginExpansion
{\displaystyle {P^\alpha P_\alpha  \over T^2}}
%EndExpansion
+X^{\prime \alpha }X_\alpha ^{\prime }.
\end{array}
\label{v.65}
\end{equation}
First of all, regarding them as canonical generators of gauge
transformations acting on string variables, we obtain\footnote{%
Since $P$ and $X$ haven't zero Poisson brackets with these constraints $\chi
_1$ and $\chi _2$, they aren't gauge-invariant at all: if one wanted to get
really gauge-invariant variables for the string, one should make one more
canonical transformation, that is one more Dirac-Bergman reduction. This
will be tried in \cite{matebiga}.}: 
\begin{equation}
\left\{ 
\begin{array}{l}
\left\{ X_\mu \left( \sigma ,\tau \right) ,\chi _1\left( \sigma ^{\prime
},\tau \right) \right\} =\partial _\sigma X_\mu \left( \sigma ,\tau \right)
\delta \left( \sigma ^{\prime }-\sigma \right) , \\ 
\\ 
\left\{ P_\mu \left( \sigma ,\tau \right) ,\chi _1\left( \sigma ^{\prime
},\tau \right) \right\} =-P_\mu \left( \sigma ,\tau \right) \partial
_{\sigma ^{\prime }}\delta \left( \sigma ^{\prime }-\sigma \right) , \\ 
\\ 
\left\{ X_\mu \left( \sigma ,\tau \right) ,\chi _2\left( \sigma ^{\prime
},\tau \right) \right\} =%
%TCIMACRO{\dfrac 2{T^2} }
%BeginExpansion
{\displaystyle {2 \over T^2}}
%EndExpansion
P_\mu \left( \sigma ,\tau \right) \delta \left( \sigma ^{\prime }-\sigma
\right) , \\ 
\\ 
\left\{ P_\mu \left( \sigma ,\tau \right) ,\chi _2\left( \sigma ^{\prime
},\tau \right) \right\} =-2\partial _\sigma X_\mu \left( \sigma ,\tau
\right) \partial _{\sigma ^{\prime }}\delta \left( \sigma ^{\prime }-\sigma
\right) .
\end{array}
\right.  \label{v.66}
\end{equation}
The only very understandable formula in (\ref{v.66}) is the first one 
\[
\left\{ X_\mu \left( \sigma ,\tau \right) ,\chi _1\left( \sigma ^{\prime
},\tau \right) \right\} =\partial _\sigma X_\mu \left( \sigma ,\tau \right)
\delta \left( \sigma ^{\prime }-\sigma \right) , 
\]
which regards $\chi _1$ as a canonical generator of translations along $%
\sigma $. In fact, defining the quantity 
\begin{equation}
D_f\left( \tau \right) =%
%TCIMACRO{\dint }
%BeginExpansion
\displaystyle \int 
%EndExpansion
f\left( \sigma ,\tau \right) \chi _1\left( \sigma ,\tau \right) d\sigma
\label{v.67}
\end{equation}
and then using it to transform canonically string variable $X_\mu \left(
\sigma ,\tau \right) $, we deduce 
\begin{equation}
\left\{ X_\mu \left( \sigma ,\tau \right) ,D_f\left( \tau \right) \right\}
=f\left( \sigma ,\tau \right) \partial _\sigma X_\mu \left( \sigma ,\tau
\right) :  \label{v.68}
\end{equation}
this is typically the action of the canonical generator of the
transformation 
\[
\begin{array}{ccc}
\tilde \tau =\tau , &  & \tilde \sigma =\sigma +f\left( \sigma ,\tau \right)
,
\end{array}
\]
always thinking of $f\left( \sigma ,\tau \right) $ as an ''infinitesimal''
function.

Such a polite relationship is rather difficult to single out for the other
equations (\ref{v.66}): less cumbersome, more encouraging results are
obtained by combining functionally the $\chi _k$'s. For example, using the
following functional 
\begin{equation}
M_f\left( \tau \right) =-%
%TCIMACRO{\dint }
%BeginExpansion
\displaystyle \int 
%EndExpansion
f\left( \sigma ,\tau \right) \left[ \frac{h^{\tau \sigma }\left( \sigma
,\tau \right) }{h^{\tau \tau }\left( \sigma ,\tau \right) }\chi _1\left(
\sigma ,\tau \right) +\frac T{2h^{\tau \tau }\left( \sigma ,\tau \right) 
\sqrt{h\left( \sigma ,\tau \right) }}\chi _2\left( \sigma ,\tau \right)
\right] d\sigma  \label{v.69}
\end{equation}
as a canonical generator, the following result is obtained 
\begin{equation}
\left\{ X_\mu \left( \sigma ,\tau \right) ,M_f\left( \tau \right) \right\}
=f\left( \sigma ,\tau \right) \dot X_\mu \left( \sigma ,\tau \right) ,
\label{v.70}
\end{equation}
while when $M_f\left( \tau \right) $ acts on the canonical momentum $P_\mu
\left( \sigma ,\tau \right) $ it yields: 
\begin{equation}
\left\{ P_\mu \left( \sigma ,\tau \right) ,M_f\left( \tau \right) \right\}
=-\partial _\sigma \left[ f\left( \sigma ,\tau \right) \left( \frac{h^{\tau
\sigma }\left( \sigma ,\tau \right) }{h^{\tau \tau }\left( \sigma ,\tau
\right) }P_\mu \left( \sigma ,\tau \right) +\frac T{2h^{\tau \tau }\left(
\sigma ,\tau \right) \sqrt{h\left( \sigma ,\tau \right) }}\partial _\sigma
X_\mu \left( \sigma ,\tau \right) \right) \right]  \label{v.71}
\end{equation}
(these are both worked out using (\ref{v.22}) and (\ref{v.23}) equations).

\subsection{Conformal gauge fixing.}

In order to render 
\[
S=-\frac T2%
%TCIMACRO{\dint }
%BeginExpansion
\displaystyle \int 
%EndExpansion
d\tau 
%TCIMACRO{\dint }
%BeginExpansion
\displaystyle \int 
%EndExpansion
d\sigma \sqrt{h}\left( h^{\tau \tau }\dot X^\mu \dot X_\mu +2h^{\tau \sigma }%
\dot X^\mu X_\mu ^{\prime }+h^{\sigma \sigma }X^{\prime \mu }X_\mu ^{\prime
}\right) 
\]
simpler, we'd like to get a diagonal sheet metric.

Let's use the following gauge fixing 
\begin{equation}
h^{{\rm ab}}-\eta ^{{\rm ab}}=0,  \label{gauge.conforme}
\end{equation}
where $\eta ^{{\rm ab}}$ is simply the Minkowskian $1+1$ metric. Since we
want to get the configuration 
\begin{equation}
\left\| h_0^{{\rm ab}}\right\| =\left( 
\begin{array}{cc}
-1 & 0 \\ 
0 & 1
\end{array}
\right)  \label{v.103}
\end{equation}
we'll have to choose a transformation (\ref{v.93}) with coefficients: 
\begin{equation}
\left\| \epsilon ^{{\rm ab}}\right\| =\left( 
\begin{array}{cc}
-1-\tilde h^{\tau \tau } & -\tilde h^{\tau \sigma } \\ 
-\tilde h^{\tau \sigma } & 1-\tilde h^{\sigma \sigma }
\end{array}
\right) .  \label{v.104}
\end{equation}

This gauge fixing is possible, since (\ref{v.104}) is always an admitted
choice; moreover, gauge fixing (\ref{gauge.conforme}) is complete, since
it's unstable under further $\pi _{{\rm ef}}$-generated transformation
because: 
\begin{equation}
\left\{ h^{{\rm ab}}\left( \sigma ,\tau \right) -\eta ^{{\rm ab}},\pi _{{\rm %
ef}}\left( \sigma ^{\prime },\tau \right) \right\} =\frac 12\left( \delta _{%
{\rm e}}^{{\rm a}}\delta _{{\rm f}}^{{\rm b}}+\delta _{{\rm f}}^{{\rm a}%
}\delta _{{\rm e}}^{{\rm b}}\right) \delta \left( \sigma ^{\prime }-\sigma
\right) \neq 0.  \label{v.105}
\end{equation}

From now on we'll always work in the {\it conformal gauge}, as the condition
(\ref{gauge.conforme}) is referred to. Polyakov's action becomes: 
\begin{equation}
S=\frac T2%
%TCIMACRO{\dint }
%BeginExpansion
\displaystyle \int 
%EndExpansion
d\tau 
%TCIMACRO{\dint }
%BeginExpansion
\displaystyle \int 
%EndExpansion
d\sigma \left( \dot X^\mu \dot X_\mu -X^{\prime \mu }X_\mu ^{\prime }\right)
.  \label{azione.conforme}
\end{equation}

We'll gauge out the gauge degrees of freedom related to the constraints 
\begin{equation}
\begin{array}{ccccc}
\pi _{\tau \tau }\approx 0, &  & \pi _{\tau \sigma }\approx 0, &  & \pi
_{\sigma \sigma }\approx 0,
\end{array}
\label{v.106}
\end{equation}
since by the conformal fixing 
\begin{equation}
\begin{array}{ccccc}
h^{\tau \tau }+1=0, &  & h^{\tau \sigma }=0, &  & h^{\sigma \sigma }-1=0
\end{array}
\label{v.107}
\end{equation}
first class set (\ref{v.106}) will be changed into a second class set,
adding the $\left( h^{{\rm ab}}-\eta ^{{\rm ab}}\right) $'s: 
\begin{equation}
\begin{array}{cccccc}
\pi _{\tau \tau }\approx 0, & \pi _{\tau \sigma }\approx 0, & \pi _{\sigma
\sigma }\approx 0, & h^{\tau \tau }+1\approx 0, & h^{\tau \sigma }\approx 0,
& h^{\sigma \sigma }-1\approx 0.
\end{array}
\label{v.108}
\end{equation}
It's possible to read as strong equations these (\ref{v.108}), using
suitable {\it Dirac brackets} instead of usual symplectic product (\ref
{poisson.1}).

The symplectic matrix of the second class constraints is defined as: 
\begin{equation}
C\left( \sigma ,\sigma ^{\prime }\right) =\left( 
\begin{array}{ccc}
\left\{ \pi _{\tau \tau }\left( \sigma \right) ,h^{\tau \tau }\left( \sigma
^{\prime }\right) +1\right\} & \left\{ \pi _{\tau \tau }\left( \sigma
\right) ,h^{\tau \sigma }\left( \sigma ^{\prime }\right) \right\} & \left\{
\pi _{\tau \tau }\left( \sigma \right) ,h^{\sigma \sigma }\left( \sigma
^{\prime }\right) -1\right\} \\ 
\left\{ \pi _{\tau \sigma }\left( \sigma \right) ,h^{\tau \tau }\left(
\sigma ^{\prime }\right) +1\right\} & \left\{ \pi _{\tau \sigma }\left(
\sigma \right) ,h^{\tau \sigma }\left( \sigma ^{\prime }\right) \right\} & 
\left\{ \pi _{\tau \sigma }\left( \sigma \right) ,h^{\sigma \sigma }\left(
\sigma ^{\prime }\right) -1\right\} \\ 
\left\{ \pi _{\sigma \sigma }\left( \sigma \right) ,h^{\tau \tau }\left(
\sigma ^{\prime }\right) +1\right\} & \left\{ \pi _{\sigma \sigma }\left(
\sigma \right) ,h^{\tau \sigma }\left( \sigma ^{\prime }\right) \right\} & 
\left\{ \pi _{\sigma \sigma }\left( \sigma \right) ,h^{\sigma \sigma }\left(
\sigma ^{\prime }\right) -1\right\}
\end{array}
\right) ,  \label{v.109}
\end{equation}
and reads: 
\begin{equation}
C\left( \sigma ,\sigma ^{\prime }\right) =-\left( 
\begin{array}{ccc}
1 & 0 & 0 \\ 
0 & 1 & 0 \\ 
0 & 0 & 1
\end{array}
\right) \delta \left( \sigma -\sigma ^{\prime }\right)  \label{matrice.c}
\end{equation}
Its inverse matrix has the same form: 
\begin{equation}
C^{-1}\left( \sigma ,\sigma ^{\prime }\right) =-\left( 
\begin{array}{ccc}
1 & 0 & 0 \\ 
0 & 1 & 0 \\ 
0 & 0 & 1
\end{array}
\right) \delta \left( \sigma -\sigma ^{\prime }\right) .  \label{matrice.c-1}
\end{equation}

Dirac brackets are defined as follows: 
\begin{equation}
\left\{ {\cal F},{\cal G}\right\} ^{*}=\left\{ {\cal F},{\cal G}\right\} -%
%TCIMACRO{\dint }
%BeginExpansion
\displaystyle \int 
%EndExpansion
d\sigma 
%TCIMACRO{\dint }
%BeginExpansion
\displaystyle \int 
%EndExpansion
d\sigma ^{\prime }\left\{ {\cal F},\Psi _A\left( \sigma \right) \right\}
\left( C^{-1}\left( \sigma ,\sigma ^{\prime }\right) \right) ^{AB}\left\{
\Psi _B\left( \sigma ^{\prime }\right) ,{\cal G}\right\} ,  \label{dirac.1}
\end{equation}
where $\Psi _A$ are (\ref{v.108}) constraints, with $A$ from $1$ to $6$;
from the form 
\begin{equation}
\left( C^{-1}\left( \sigma ,\sigma ^{\prime }\right) \right) ^{AB}=-\delta
^{AB}\delta \left( \sigma -\sigma ^{\prime }\right) ,  \label{v.110}
\end{equation}
we immediately get: 
\begin{equation}
\left\{ {\cal F},{\cal G}\right\} ^{*}=\left\{ {\cal F},{\cal G}\right\} +%
%TCIMACRO{\dsum }
%BeginExpansion
\mathop{\displaystyle \sum }
%EndExpansion
\limits_{A=1}^6%
%TCIMACRO{\dint }
%BeginExpansion
\displaystyle \int 
%EndExpansion
d\sigma \left\{ {\cal F},\Psi _A\left( \sigma \right) \right\} \left\{ \Psi
_A\left( \sigma \right) ,{\cal G}\right\} .  \label{dirac.2}
\end{equation}
Explicitly: 
\begin{equation}
\begin{array}{l}
\left\{ {\cal F},{\cal G}\right\} ^{*}=\left\{ {\cal F},{\cal G}\right\} +
\\ 
\\ 
+%
%TCIMACRO{\dint }
%BeginExpansion
\displaystyle \int 
%EndExpansion
d\sigma \left\{ {\cal F},\pi _{\tau \tau }\left( \sigma \right) \right\}
\left\{ \pi _{\tau \tau }\left( \sigma \right) ,{\cal G}\right\} +%
%TCIMACRO{\dint }
%BeginExpansion
\displaystyle \int 
%EndExpansion
d\sigma \left\{ {\cal F},\pi _{\tau \sigma }\left( \sigma \right) \right\}
\left\{ \pi _{\tau \sigma }\left( \sigma \right) ,{\cal G}\right\} + \\ 
\\ 
+%
%TCIMACRO{\dint }
%BeginExpansion
\displaystyle \int 
%EndExpansion
d\sigma \left\{ {\cal F},\pi _{\sigma \sigma }\left( \sigma \right) \right\}
\left\{ \pi _{\sigma \sigma }\left( \sigma \right) ,{\cal G}\right\} +%
%TCIMACRO{\dint }
%BeginExpansion
\displaystyle \int 
%EndExpansion
d\sigma \left\{ {\cal F},h^{\tau \tau }\left( \sigma \right) \right\}
\left\{ h^{\tau \tau }\left( \sigma \right) ,{\cal G}\right\} \\ 
\\ 
+%
%TCIMACRO{\dint }
%BeginExpansion
\displaystyle \int 
%EndExpansion
d\sigma \left\{ {\cal F},h^{\tau \sigma }\left( \sigma \right) \right\}
\left\{ h^{\tau \sigma }\left( \sigma \right) ,{\cal G}\right\} +%
%TCIMACRO{\dint }
%BeginExpansion
\displaystyle \int 
%EndExpansion
d\sigma \left\{ {\cal F},h^{\sigma \sigma }\left( \sigma \right) \right\}
\left\{ h^{\sigma \sigma }\left( \sigma \right) ,{\cal G}\right\} .
\end{array}
\label{dirac.3}
\end{equation}

Since Dirac bracketing allows us to write 
\[
\Psi _A\left( \sigma ,\tau \right) =0 
\]
strongly, we can directly modify quantities involving the gauged out
constraints. From string velocity to string momentum now we go by: 
\begin{equation}
P_\mu =T\dot X_\mu  \label{v.111}
\end{equation}
and come back by: 
\begin{equation}
\dot X_\mu =\frac{P_\mu }T,  \label{v.112}
\end{equation}
which replace (\ref{v.22}) and (\ref{v.23}) respectively.

Linear canonical Hamiltonian density becomes: 
\begin{equation}
{\cal H}_{{\rm C}}=%
%TCIMACRO{\dfrac{P^\alpha P_\alpha }{2T} }
%BeginExpansion
{\displaystyle {P^\alpha P_\alpha  \over 2T}}
%EndExpansion
+%
%TCIMACRO{\dfrac T2 }
%BeginExpansion
{\displaystyle {T \over 2}}
%EndExpansion
X^{\prime \alpha }X_\alpha ^{\prime },  \label{v.113}
\end{equation}
and the extended correspondent: 
\begin{equation}
{\cal H}_{{\rm E}}=\lambda _1P^\alpha X_\alpha ^{\prime }+\left( \lambda _2+%
%TCIMACRO{\dfrac T2 }
%BeginExpansion
{\displaystyle {T \over 2}}
%EndExpansion
\right) \left( 
%TCIMACRO{\dfrac{P^\alpha P_\alpha }{T^2} }
%BeginExpansion
{\displaystyle {P^\alpha P_\alpha  \over T^2}}
%EndExpansion
+X^{\prime \alpha }X_\alpha ^{\prime }\right) .  \label{v.114}
\end{equation}
Surviving constraints involving $P$ and $X$ are left formally invariant,
while their expressions in terms of $\dot X$ and\ $X$ change as: 
\begin{equation}
\begin{array}{cc}
\dot X^\alpha X_\alpha ^{\prime }\approx 0, & \dot X^\alpha \dot X_\alpha
+X^{\prime \alpha }X_\alpha ^{\prime }\approx 0.
\end{array}
\label{v.115}
\end{equation}

Dirac symplectic product among string variables reads: 
\begin{equation}
\left\{ X^\mu \left( \sigma ,\tau \right) ,\dot X_\nu \left( \sigma ^{\prime
},\tau \right) \right\} ^{*}=T\eta _\nu ^\mu \delta \left( \sigma -\sigma
^{\prime }\right) .  \label{v.116}
\end{equation}
From now on, we'll write this relationship without the asterisk $^{*}$
(which was there to remind us that a gauge reduction took place).

The equation of motion (\ref{v.79}) due to the canonical Hamiltonian is
modified as 
\begin{equation}
\dot P_\mu =TX_\mu ^{\prime \prime },  \label{v.117}
\end{equation}
that is: 
\begin{equation}
\ddot X_\mu -X_\mu ^{\prime \prime }=0.  \label{eqonda.3}
\end{equation}

Working with the extended Hamiltonian (\ref{v.114}) we would get 
\begin{equation}
\ddot X_\mu -\lambda _1^{\prime }\dot X_\mu -\lambda _1\dot X_\mu ^{\prime
}-2\frac{\lambda _2^{\prime }}TX_\mu ^{\prime }-\left( 2\frac{\lambda _2}T%
+1\right) X_\mu ^{\prime \prime }=0.  \label{eqonda.4}
\end{equation}

After deciding the conformal gauge (\ref{gauge.conforme}) to be used from
here on, let's show that its conclusions are coherent with what's been
discovered within Lagrangean framework. There (in (\ref{v.4})) it was
stressed that Euler-Lagrange equations for $h^{{\rm ab}}$ led to: 
\begin{equation}
h_{{\rm bc}}=\eta _{\mu \nu }\partial _{{\rm b}}X^\mu \partial _{{\rm c}%
}X^\nu ,  \label{embedh.1}
\end{equation}
i.e. the sheet metric from Lagrangean equations was shown to be that of $%
{\Bbb V}_2$ when embedded into $M_D$. This spacetime is flat, anyway, and
has $\eta _{\mu \nu }$ as metric tensor, so we have: 
\begin{equation}
\begin{array}{ccc}
h_{\tau \tau }=\dot X^2, & h_{\sigma \sigma }=\left( X^{\prime }\right) ^2,
& h_{\tau \sigma }=\dot X\cdot X^{\prime }.
\end{array}
\label{embedh.2}
\end{equation}
Constraints (\ref{v.115}) simply tell us that the choice 
\[
\begin{array}{ccc}
h_{\tau \tau }=-1, & h_{\sigma \sigma }=+1, & h_{\tau \sigma }=0
\end{array}
\]
is admitted, consistently with (\ref{embedh.1}) and (\ref{embedh.2}).

\subsection{Surviving gauge symmetries.}

Let's deal with the string theory in which the sheet constraints $\pi _{{\rm %
ab}}\approx 0$ have been gauged away by the position (\ref{gauge.conforme}):
now we've not any more five constraints, but only two first class ones (the
weakly vanishing $\chi _1$ and $\chi _2$).

This new theory still shows all the symmetries generated by the two
surviving constraints: 
\begin{equation}
\begin{array}{cc}
P^\mu X_\mu ^{\prime }\approx 0, & 
%TCIMACRO{\dfrac{P^\alpha P_\alpha }{T^2} }
%BeginExpansion
{\displaystyle {P^\alpha P_\alpha  \over T^2}}
%EndExpansion
+X^{\prime \alpha }X_\alpha ^{\prime }\approx 0.
\end{array}
\label{v.118}
\end{equation}
Constraints (\ref{v.118}) generate canonically all the worldsheet
transformations which leave the conformal action 
\begin{equation}
S=\frac T2%
%TCIMACRO{\dint }
%BeginExpansion
\displaystyle \int 
%EndExpansion
d\tau 
%TCIMACRO{\dint }
%BeginExpansion
\displaystyle \int 
%EndExpansion
d\sigma \left[ \dot X^2-\left( X^{\prime }\right) ^2\right]  \label{v.119}
\end{equation}
unchanged.

In this paragraph we want to show that those gauge transformations which
survive after the gauge-fixing (\ref{gauge.conforme}) can be rewritten
producing the very well known Virasoro algebra: due to this fact, string
theory will be interpreted as a conformal field theory, that's one of its
most pregnant property.

\subsubsection{Stress-energy tensor.}

The first thing we want to stress is the meaning of the quantities $\chi _1$
and $\chi _2$ in the conformal gauge we've chosen; they correspond to the
stress-energy tensor components, defined as follows \cite{Carroll}: 
\begin{equation}
T_{{\rm ab}}\left( \eta \right) =-\left. \frac 2{T\sqrt{h}}\frac{\delta S}{%
\delta h^{{\rm ab}}}\right| _{h=\eta }.  \label{v.120}
\end{equation}
In order to evaluate this $T_{{\rm ab}}\left( \eta \right) $ it's necessary
to use the still gauge-unfixed Polyakov's action (\ref{azione.pol}), compute
the derivatives of (\ref{v.120}), and finally restrict ourselves to the
conformal choice for the gauge. The tensor is evaluated as 
\begin{equation}
T_{{\rm ab}}\left( h\right) =\partial _{{\rm a}}X^\mu \partial _{{\rm b}%
}X_\mu -\frac 12h_{{\rm ab}}\left( h^{{\rm ef}}\partial _{{\rm e}}X^\mu
\partial _{{\rm f}}X_\mu \right) .  \label{v.122}
\end{equation}
In the conformal gauge the latter becomes: 
\begin{equation}
T_{{\rm ab}}\left( \eta \right) =\partial _{{\rm a}}X^\mu \partial _{{\rm b}%
}X_\mu +\frac 12\eta _{{\rm ab}}\left( \dot X^2-\left( X^{\prime }\right)
^2\right) .  \label{v.123}
\end{equation}
When conformally gauge-fixed, the stress-energy tensor elements are the very
constraints $\chi _1$ and $\chi _2$ which survive the gauge-fixing as
further symmetries: 
\begin{equation}
\begin{array}{ccc}
T_{\tau \tau }\left( \eta \right) =%
%TCIMACRO{\dfrac 12 }
%BeginExpansion
{\displaystyle {1 \over 2}}
%EndExpansion
\left( \dot X^2+\left( X^{\prime }\right) ^2\right) , & T_{\sigma \sigma
}\left( \eta \right) =%
%TCIMACRO{\dfrac 12 }
%BeginExpansion
{\displaystyle {1 \over 2}}
%EndExpansion
\left( \dot X^2+\left( X^{\prime }\right) ^2\right) , & T_{\tau \sigma
}\left( \eta \right) =\dot X\cdot X^{\prime },
\end{array}
\label{v.124}
\end{equation}
that is 
\begin{equation}
\begin{array}{cc}
T_{\tau \sigma }\left( \eta \right) =\chi _1, & T_{\tau \tau }\left( \eta
\right) =T_{\sigma \sigma }\left( \eta \right) =%
%TCIMACRO{\dfrac 12 }
%BeginExpansion
{\displaystyle {1 \over 2}}
%EndExpansion
\chi _2.
\end{array}
\label{v.124.bis}
\end{equation}
So within the conformal gauge the theory shows the following constraints: 
\begin{equation}
T_{{\rm ab}}\approx 0,  \label{v.125}
\end{equation}
omitting the gauge-fixing symbol $\left( \eta \right) $.

This tensor $T_{{\rm ab}}$ is interesting, because it corresponds in general
to the sheet stress-energy tensor, that is to $\left( 1+1\right) $-current
which describes the flux of canonical $\sigma $- and $\tau $-translation
generators along the worldsheet, during the motion of $X^\mu \left( \sigma
,\tau \right) $. This is deduced from the application of Noether's theorem
to Polyakov's Lagrangean density.

The sheet densities of the canonical generators of translations along $\tau $
and $\sigma $ are 
\begin{equation}
\rho _{{\rm a}}=-\dot X^\mu \partial _{{\rm a}}X_\mu +\frac 12\eta ^\tau \,_{%
{\rm a}}\left( \dot X^2-\left( X^{\prime }\right) ^2\right) ,  \label{v.129}
\end{equation}
where ${\rm a}=\tau ,\,\sigma $, and the current density of the {\rm a}-th
canonical generator simply is: 
\begin{equation}
J_{{\rm a}}=-X^{\prime \mu }\partial _{{\rm a}}X_\mu +\frac 12\eta ^\sigma
\,_{{\rm a}}\left( \dot X^2-\left( X^{\prime }\right) ^2\right) .
\label{v.130}
\end{equation}
It's easy to show that the equation 
\begin{equation}
\dot \rho _{{\rm a}}+J_{{\rm a}}^{\prime }=\partial _{{\rm a}}X^\mu \left( 
\ddot X_\mu -X_\mu ^{\prime \prime }\right)  \label{v.131}
\end{equation}
is fulfilled by $\rho _{{\rm a}}$ and $J_{{\rm a}}$, which becomes a
continuity equation $\dot \rho _{{\rm a}}+J_{{\rm a}}^{\prime }=0$ when
equations of motion (\ref{eqonda.3}) are considered. So one can write 
\begin{equation}
\dot \rho _{{\rm a}}+J_{{\rm a}}^{\prime }\stackrel{\circ }{=}0.
\label{v.132}
\end{equation}
The canonical generators of sheet translations are obtained from $\rho _\tau 
$ and $\rho _\sigma $ integrating along the string at fixed $\tau $, and
are: 
\begin{equation}
\begin{array}{cc}
Q_\tau =%
%TCIMACRO{\dint }
%BeginExpansion
\displaystyle \int 
%EndExpansion
d\sigma T^\tau \,_\tau \left( \sigma ,\tau \right) , & Q_\sigma =%
%TCIMACRO{\dint }
%BeginExpansion
\displaystyle \int 
%EndExpansion
d\sigma T^\tau \,_\sigma \left( \sigma ,\tau \right) .
\end{array}
\label{v.133}
\end{equation}
Continuity laws (\ref{v.132}) allow the relationships 
\begin{equation}
\begin{array}{cc}
\dot Q_\tau =0, & \dot Q_\sigma =0
\end{array}
\label{v.134}
\end{equation}
to hold along the string motion.

Let's turn for a moment to equations (\ref{v.124}): from them one can derive 
\begin{equation}
\begin{array}{cc}
\rho _\tau =-%
%TCIMACRO{\dfrac 12 }
%BeginExpansion
{\displaystyle {1 \over 2}}
%EndExpansion
\chi _2, & \rho _\sigma =\chi _1;
\end{array}
\label{v.135}
\end{equation}
this makes $\rho _\tau $ and $\rho _\sigma $ have zero Poisson brackets.
Moreover, since the canonical Hamiltonian is (see equation (\ref{v.113})) 
\begin{equation}
{\cal H}_{{\rm C}}=\frac T2\chi _2  \label{v.136}
\end{equation}
it's to be expected that $Q_\tau $ and $Q_\sigma $ are constant along the
motion.

The matrix of the sheet stress-energy tensor $T_{{\rm ab}}$ contains the
constraints $\chi _k$'s: 
\[
\left\| T_{{\rm ab}}\right\| =\left( 
\begin{array}{cc}
\frac 12\chi _2 & \chi _1 \\ 
\chi _1 & \frac 12\chi _2
\end{array}
\right) . 
\]
The weak vanishing of $T_{{\rm ab}}$ components is completely equivalent to
that of $\chi _k$'s, while their linear combination 
\begin{equation}
T_{{\rm ab}}h^{{\rm ba}}\approx 0  \label{v.138}
\end{equation}
vanishes as a stability condition for Weyl's rescaling invariance of
Polyakov's action, as it is shown by equation (\ref{traccia.4}).

\subsubsection{Virasoro algebra.}

There's a local conservation law 
\begin{equation}
\partial ^{{\rm a}}T_{{\rm ab}}=0  \label{continuita}
\end{equation}
to which tensor $T_{{\rm ab}}$ undergoes, and it allows us to build up an
infinite number of conserved quantities for Polyakov's string motion. All
these quantities form the Poisson algebra of the rich group of {\it %
conformal transformations}, the very gauge symmetries generated by $\chi _k$%
's.

First of all, we have to extend string variables between $\sigma =-\pi $ and 
$\sigma =0$ in order to produce those conserved quantities; an even
extension is introduced for the velocity 
\begin{equation}
\dot X_\mu \left( -\sigma ,\tau \right) =\dot X_\mu \left( \sigma ,\tau
\right) ,  \label{v.139}
\end{equation}
while gradients must undergo to an odd extension: 
\begin{equation}
X_\mu ^{\prime }\left( -\sigma ,\tau \right) =-X_\mu ^{\prime }\left( \sigma
,\tau \right) .  \label{v.140}
\end{equation}

Let's then consider a smooth function 
\[
f\in C^\infty \left( \left[ -\pi ,\pi \right] ,{\Bbb C}\right) 
\]
and define the following functional of secondary constraints: 
\begin{equation}
L\left[ f\right] =\frac T4%
%TCIMACRO{\dint }
%BeginExpansion
\displaystyle \int 
%EndExpansion
\limits_{-\pi }^{+\pi }f\left( \sigma \right) \left[ \frac{2\chi _1\left(
\sigma \right) }T+\chi _2\left( \sigma \right) \right] d\sigma ,
\label{virasoro.1}
\end{equation}
which reads 
\begin{equation}
L\left[ f\right] =\frac T4\eta ^{\mu \nu }%
%TCIMACRO{\dint }
%BeginExpansion
\displaystyle \int 
%EndExpansion
\limits_{-\pi }^{+\pi }f\left( \sigma \right) \left[ \frac{P_\mu \left(
\sigma \right) }T+X_\mu ^{\prime }\left( \sigma \right) \right] \left[ \frac{%
P_\nu \left( \sigma \right) }T+X_\nu ^{\prime }\left( \sigma \right) \right]
d\sigma  \label{virasoro.2}
\end{equation}
in terms of string variables.

The important fact is that these functionals have a closed Poisson bracket
algebra. The linear combination of functionals expressing the Poisson
bracket of two given functionals $L\left[ f\right] $ and $L\left[ g\right] $
is rather easy: 
\[
\left\{ L\left[ f\right] ,L\left[ g\right] \right\} =\frac T4%
%TCIMACRO{\dint}
%BeginExpansion
\displaystyle \int 
%EndExpansion
\limits_{-\pi }^{+\pi }\left[ f\left( \sigma \right) g^{\prime }\left(
\sigma \right) -f^{\prime }\left( \sigma \right) g\left( \sigma \right)
\right] \left[ \frac{2\chi _1\left( \sigma \right) }T+\chi _2\left( \sigma
\right) \right] d\sigma , 
\]
i.e. by definition (\ref{virasoro.1}): 
\begin{equation}
\left\{ L\left[ f\right] ,L\left[ g\right] \right\} =L\left[ fg^{\prime
}-f^{\prime }g\right] .  \label{virasoro.3}
\end{equation}

Defining: 
\begin{equation}
fg^{\prime }-f^{\prime }g=f\times g  \label{v.141}
\end{equation}
one has: 
\begin{equation}
\left\{ L\left[ f\right] ,L\left[ g\right] \right\} =L\left[ f\times
g\right] .  \label{virasoro.4}
\end{equation}
This tells us that equation (\ref{virasoro.1}) functionals span a closed
symplectic algebra \cite{Kaku}. This gives a canonical realization of the
conformal group \cite{pauri-prosperi.1}. This canonical group is very big,
so big as the set of functions $f$ and $g$ which is possible to construct
the $L$'s with, so it's worth simplifying it ordering its elements in a more
transparent way. We can do this Fourier-decomposing functions $f$ as: 
\begin{equation}
f\left( \sigma \right) =%
%TCIMACRO{\dsum }
%BeginExpansion
\mathop{\displaystyle \sum }
%EndExpansion
\limits_{n\in {\Bbb Z}}A_n\exp \left( in\sigma \right) ,  \label{v.142}
\end{equation}
This induces the following basis for the algebra (\ref{virasoro.4}) 
\begin{equation}
L_n=\frac T4%
%TCIMACRO{\dint }
%BeginExpansion
\displaystyle \int 
%EndExpansion
\limits_{-\pi }^{+\pi }d\sigma \exp \left( in\sigma \right) \left[ \frac{%
2\chi _1\left( \sigma \right) }T+\chi _2\left( \sigma \right) \right] .
\label{virasoro.5}
\end{equation}

These $L_n$ are the very well known {\it Virasoro charges}: it's evident
that the conservation of these theoretical charges 
\begin{equation}
\dot L_n=0  \label{virasoro.6}
\end{equation}
follows without problems from secondary constraint stability. Equation (\ref
{virasoro.6}) yields as well 
\begin{equation}
\begin{array}{ccc}
\dot L\left[ f\right] =0 & \forall & f,
\end{array}
\label{virasoro.7}
\end{equation}
by the definition of (\ref{virasoro.1}).

In terms of ${\Bbb Z}$-number indices the algebra(\ref{virasoro.4}) becomes: 
\begin{equation}
\left\{ L_m,L_n\right\} =-i\left( m-n\right) L_{m+n},  \label{virasoro.8}
\end{equation}
which is referred to as {\it Virasoro classical algebra} \cite{Fuchs}. By
definition, Virasoro charges are thus the very constraints of conformal
invariance: 
\begin{equation}
L_m\approx 0.  \label{virasoro.9}
\end{equation}

\acknowledgments

I am very grateful to Prof. L. Lusanna of Firenze University for having
suggested me the consultation of references \cite{Teitelboim}, \cite
{marnelius} and \cite{pronko}.

I would like to thank Doctor C. Palmisano too, for reading the manuscript.

\end{document}